\documentclass[%
twocolumn,
 floatfix,
 superscriptaddress,
nofootinbib,
 amsmath,amssymb,
 aps,
]{revtex4-2}

\usepackage{booktabs}%making tables
\usepackage[x11names]{xcolor}
\usepackage{graphicx}% Include figure files
\usepackage{dcolumn}% equation table columns on decimal point
\usepackage{bm}% bold math
\usepackage{orcidlink}
\usepackage{hyperref}% add hypertext capabilities
\hypersetup{
    colorlinks=true,
    linkcolor=red,
    citecolor=blue,
    filecolor=magenta,
    urlcolor=blue
}
 \usepackage{ragged2e}
\usepackage{cleveref}% clever references in latex
\usepackage[mathlines]{lineno}% Enable numbering of text and display math
\usepackage{todonotes}

\usepackage[normalem]{ulem}
\usepackage[caption=false,position=top]{subfig}
\usepackage[x11names,dvipsnames,svgnames]{xcolor}
\usepackage{custom}

\usepackage{comment}

\newcommand{\sNN}{\sqrt{s_{\rm NN}}}
\newcommand{\low}{_\text{low}}
\newcommand{\mus}{\mu_B,\mu_Q,\mu_S}
\newcommand{\boundary}{\bigg|_{T_{0,\text{low}}}^{T_0}}

\newcommand{\dth}[1]{\left(\frac{\partial #1}{\partial\theta}\right)_{\tilde T_0,\varphi}}
\newcommand{\dph}[1]{\left(\frac{\partial #1}{\partial\varphi}\right)_{\tilde T_0,\theta}}

\allowdisplaybreaks
\begin{document}

\title{
Four-dimensional QCD equation of state with a critical surface from contours of constant entropy density
}

\author{Tristan Gyure}
\affiliation{
 Department of Physics, University of Houston, Houston, TX 77204, USA
}

\author{Hitansh Shah\,\orcidlink{0009-0008-1870-3157}}
\affiliation{
 Department of Physics, University of Houston, Houston, TX 77204, USA
}
\affiliation{Institute for Theoretical Physics, Goethe University,
Max-von-Laue-Strasse 1, 60438 Frankfurt am Main, Germany}

\author{Lorenzo Formaggio}
\affiliation{
 Department of Physics, University of Houston, Houston, TX 77204, USA
}

\author{Francesco Di Clemente\,\orcidlink{0000-0002-8257-3819}}
\affiliation{
 Department of Physics, University of Houston, Houston, TX 77204, USA
}

\author{Mauricio Hippert\,\orcidlink{0000-0001-5802-3908}}
% \email{hippert@cbpf.br}
\affiliation{Centro Brasileiro de Pesquisas F\'isicas, Rua Dr. Xavier Sigaud 150, Rio de Janeiro, RJ, 22290-180, Brazil}

\author{Claudia Ratti\,\orcidlink{0000-0002-8335-567X}}
\affiliation{
 Department of Physics, University of Houston, Houston, TX 77204, USA
}

\author{Volodymyr Vovchenko\,\orcidlink{0000-0002-2189-4766}}
\affiliation{
 Department of Physics, University of Houston, Houston, TX 77204, USA
}

\date{\today}

\begin{abstract}

We present a new lattice-QCD-based equation of state in the four-dimensional space of $(T,\mus)$ which features a critical surface. 
This is obtained by generalizing the expansion of constant entropy density contours across the multiple conserved charges of QCD. 
Since this expansion permits the presence of a first-order phase transition, this is the first lattice QCD based equation of state in four dimensions with a critical surface.
We parametrize the chemical potential space using spherical coordinates $(\mu,\theta,\varphi)$, reducing the problem to a one-dimensional extrapolation in $(T,\mu)$ at fixed angles that, together, span the full 4D space. By integrating the extrapolated entropy density, we reconstruct the 
pressure, from which we obtain the conserved charge densities, their susceptibilities and the speed of sound.  Our results agree quantitatively with other lattice QCD constructions at low chemical potentials in the crossover region, and predict the emergence of a first-order phase transition at higher chemical potentials. 
We study equation of state properties in detail under 
heavy-ion collision (HIC) conditions for the strangeness and electric charge densities ($n_S=0,~n_Q=0.4n_B$), where we find the CP location at $(T_c,\mu_{B,c},\mu_{Q,c},\mu_{S,c})=(115\pm7,646\pm72,-12\pm3,117\pm16)$ MeV. 
In particular, we find a strong non-monotonic dependence of baryon number susceptibilities along the chemical freeze-out line at $\sNN \lesssim 7.7$~GeV.
Finally, we outline further applications of the 4D equation of state, including charge-neutral matter relevant to neutron stars and large isospin density relevant to cosmic trajectories in the early Universe.
\end{abstract}

\maketitle

\section{Introduction}
\label{sec:introduction}
Understanding the different phases of strongly interacting matter described by Quantum Chromodynamics (QCD) is one of the central goals in nuclear physics. At vanishing chemical potential, lattice QCD calculations predict an analytic crossover from the hadronic gas to a deconfined quark-gluon plasma at a temperature of approximately $T\simeq155-158$ MeV \cite{Aoki:2006br,Borsanyi:2020fev,HotQCD:2018pds}. At finite chemical potentials, however, first-principles calculations are hindered by the fermion sign problem. To circumvent this, several extrapolations of lattice QCD simulations are performed, which imply that the analytic crossover persists up to baryon chemical potentials of $\mu_B/T \leq 3.5$~\cite{Borsanyi:2020fev, Vovchenko:2017gkg, Bollweg:2022rps}. 
It was hinted in Ref.~\cite{Borsanyi:2024xrx} that the crossover narrows at larger $\mu_B$.
Whether this crossover ends at a potential critical point (CP) is still an open question.  
Several approaches, including functional methods, effective models, and lattice-based extrapolations do predict the existence of a CP with an associated first-order phase transition from hadrons to deconfined quark gluon plasma around $\mu_B \approx 600$ MeV~\cite{Shah:2024img,Hippert:2023bel,Gao:2020fbl,Lu:2025cls,Grefa:2021qvt,Gunkel:2021oya,Basar:2023nkp,Clarke:2024ugt}.
QCD matter at finite temperature and density is found in relativistic heavy-ion collisions and in astrophysical phenomena such as binary neutron stars and their mergers. Heavy-ion collision experiments are currently performed at the Large Hadron collider (LHC). 
The second phase of the Beam Energy Scan (BESII) at the Relativistic Heavy-ion Collider (RHIC)~\cite{Bzdak:2019pkr,Du:2024wjm} concluded its data-taking in early 2026, while the construction of a new Facility for Antiproton and Ion Research (FAIR) is nearing completion and will continue the search for the conjectured CP~\cite{Luo:2015doi, Durante:2019hzd,Odyniec:2019kfh, Dexheimer:2020zzs,Almaalol:2022xwv}. 
BES-I and BES-II results are consistent with the absence of the CP at collider energies~($\mu_B \lesssim 400$~MeV) and indicate deviations from the non-critical baselines for net-proton cumulants at low beam energies of $\sqrt{s_{NN}} \lesssim 10$ GeV~\cite{STAR:2020tga,STAR:2021iop,Braun-Munzinger:2020jbk,Vovchenko:2021kxx}. 
Deviations at low energies may emerge in potentially relevant observables such as mean transverse momentum $p_T$ fluctuations, directed flow and other observables~\cite{STAR:2026vjv,STAR:2014clz,STAR:2014nuj,Monnai:2016kud,Plumberg:2017tvu}.

Hydrodynamic simulations play a key role in the theoretical interpretation of heavy-ion measurements.
To model the space-time evolution of QCD matter  near the QCD CP, standard relativistic viscous hydrodynamics
requires important modifications~\cite{Stephanov:2017ghc,Rajagopal:2019xwg, Du:2020bxp, Bluhm:2020mpc, An:2021wof}. 
The equation of state (EoS) with a CP is essential for incorporating the effects of criticality on fluctuations and non-trivial transport phenomena~\cite{An:2021wof,Parotto:2018pwx,Kahangirwe:2024cny,Kapusta:2022pny,Karthein:2021nxe}.
While many hydrodynamic simulations employ an EoS restricted to $(T,\mu_B)$, with strangeness neutrality and a fixed charge-to-baryon ratio imposed implicitly, this reduction holds only on average. Simulations that track local fluctuations of all conserved charges, e.g.\ from initial-state quark--antiquark production, or that include coupled diffusion of baryon number, strangeness, and electric charge, sample regions away from these constraints and therefore require the EoS over the full $(T,\mu_B,\mu_Q,\mu_S)$ space~\cite{Carzon:2019qja,Martinez:2019jbu,Jahan:2026hvs}.
\begin{figure*}[t]

    \centering
    \includegraphics[width=\textwidth]{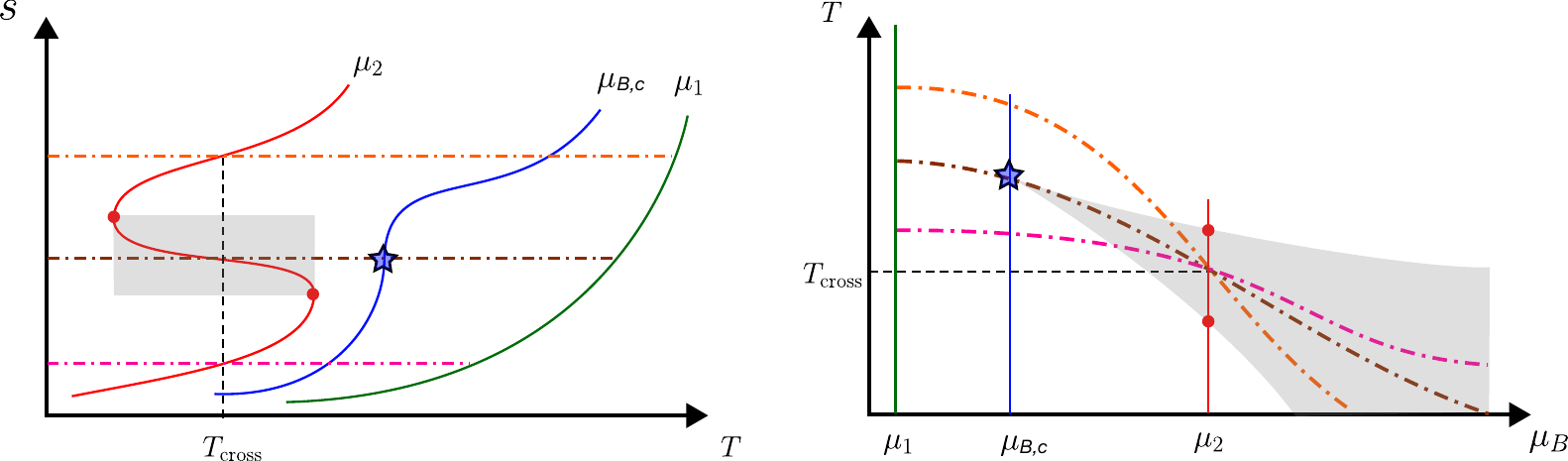}
    % \vspace{-1.5cm}
    \caption{ \justifying \small 
\textit{Left panel:} Entropy as a function of temperature for three representative baryon chemical potentials, with $\mu_1 < \mu_{B,c} < \mu_2$. 
\textit{Right panel:} The corresponding constant-entropy trajectories in the $(T, \mu_B)$ plane. 
The blue star marks the location of the CP, the shaded region indicates the spinodal domain, and the red dots highlight the spinodal points at $\mu_B = \mu_2$.  Figure taken from Ref. \cite{Shah:2024img}.
    }
    \label{fig:cartoon_firstorder}
\end{figure*}
Several effective approaches provide an EoS with a CP, such as the holographic black hole
model~\cite{Grefa:2021qvt,Knaute:2017opk, Critelli:2017oub,Jokela:2026jkh,Li:2026lxx,Chen:2024mmd}, as well as functional QCD methods~\cite{Lu:2025cls,Gunkel:2021oya,Gao:2020fbl} which can help with hydrodynamic simulations, nevertheless with their own caveats. 
Most lattice-based finite-density EoS constructions rely on Taylor expansions of the pressure around vanishing chemical potentials \cite{Bazavov:2017dus,Bollweg:2022rps,Noronha-Hostler:2019ayj,Monnai:2019hkn,Monnai:2024pvy}.
However, the applicability of the Taylor expansion is generally limited to $\mu_B/T \lesssim 3$~\cite{Vovchenko:2017gkg,Adam:2025phc}. 
To mitigate this limitation, an alternate expansion was developed by the Wuppertal Budapest lattice group, known as the $T'-$expansion scheme, which, in the pure $\mu_B$ direction, extends the region of applicability up to $\mu_B/T = 3.5$~\cite{Borsanyi:2021sxv}.
This expansion was recently extended to the full four-dimensional space of chemical potentials~\cite{Abuali:2025tbd}. 
However, neither the Taylor nor the $ T'$-expansion incorporate a CP in its EoS. 

Recently, we proposed a novel method based on the expansion of lattice QCD thermodynamics that follows contours of constant entropy density~\cite{Shah:2024img}.
This is the first lattice-based expansion method that does not break down at the CP, but rather predicts one at
$(T,\mu_B) \rightarrow (114 \pm 7,602 \pm 62)$ MeV in the pure $\mu_B$ direction~($\mu_Q = \mu_S = 0$) when truncated at $\mathcal{O}(\mu_B^2)$. This allows the CP to be incorporated into the equation of state, albeit with a mean-field critical behavior.
The method was then expanded to construct an equation of state with a line of first-order phase transition~\cite{Shah:2026pue}. 
The expansion accurately reproduces the entropy contours and CP for the holographic black hole model~\cite{Hippert:2023bel} and functional QCD approach~\cite{Lu:2025cls} at $\mu_B \approx 600$ MeV, but yields a spurious CP for the hadron resonance gas model at $\mu_B \approx 850$ MeV. 
The approach was applied by the Wuppertal-Budapest collaboration to exclude the CP for $\mu_B \leq 450$ MeV under strangeness-neutral conditions~\cite{Borsanyi:2025dyp}. 
Subsequently, in Ref.~\cite{Shah:2026qcs}, some of us generalized the method to the four-dimensional space of thermodynamic variables in (2+1)-flavor QCD, namely $(T,\mu_B,\mu_Q,\mu_S)$. 
This led to the first lattice-based prediction of a critical surface with all three conserved-charge chemical potentials. 
In this work, we construct a four-dimensional lattice-QCD-based equation of state in $(T,\mu_B,\mu_Q,\mu_S)$ with an emergent critical surface, using the generalized expansion of constant entropy density contours introduced in Ref.~\cite{Shah:2026qcs}. In this reference, we had studied the critical surface, in particular the movement of the CP's position throughout the space of multiple chemical potentials, without computing the EoS itself.
Here, we first compare the EoS obtained with the constant entropy contours method to those obtained with other expansion schemes in the case of strangeness neutrality (SN), $n_S=0$ with $\mu_Q=0$, to probe their agreement when including the strangeness chemical potential.
We then impose heavy-ion-collision (HIC) conditions, $n_S=0$ and $n_Q/n_B=0.4$, and locate the critical endpoint at $(T_c,\mu_{B,c})=(115\pm7,\,646\pm72)$~MeV. Under these conditions we study the various thermodynamic quantities such as baryon number cumulants, isentropic trajectories, and the speed of sound squared. We find a lensing effect of the isentropic trajectories in the vicinity of the CP \cite{Dore:2022qyz}, as well as the diverging and dipping of baryon cumulants and speed of sound squared, respectively. These quantities are also calculated on the chemical freeze-out line to examine the effect of the CP at hadronic freeze-out. Finally, we examine other applications of the four-dimensional equation of state, applying it to early Universe and neutron star conditions. 

The manuscript is organized as follows. In Section~\ref{sec:contours_methodology}, we review the constant entropy density contour method in two dimensions and present its extension to the full four-dimensional chemical potential space. Section~\ref{sec:eos} describes the construction of the equation of state from the entropy density at finite $(T,\mu_B,\mu_Q,\mu_S)$ obtained from the expansion. We present results for the four-dimensional equation of state in Section~\ref{sec:results}. We examine the heavy-ion collision plane in Subsec.~\ref{sec:hic_results}, while applications of the 4D-EoS to early-Universe and compact star conditions are discussed in Subsec.~\ref{sec:astro}. 
Finally, we summarize our conclusions in Section~\ref{sec:conclusions}.

\section{Contours of Constant Entropy Density Method}
\label{sec:contours_methodology}

In this section we first review the contours of constant entropy density formalism in two dimensions. We then follow the procedure used in Ref. \cite{Shah:2026qcs} to extend the method up to four dimensions, including the electric charge and strangeness chemical potentials. Once the four-dimensional formalism has been established, we discuss the lattice susceptibilities that we employ to anchor our expansion from $\mu=0$.

\subsection{Methodology in two dimensions}

The constant entropy density contour formalism was introduced in Ref.~\cite{Shah:2024img} to locate the QCD CP by extrapolating lattice QCD results for the contours of constant entropy to finite $\mu_B$. We review this method in this section. The idea is that the entropy density $s$ becomes a multi-valued function in $T$ and $\mu_B$ near a first-order phase transition, as illustrated in the left panel of Fig.~\ref{fig:cartoon_firstorder}. Hence, by observing when contours of constant entropy density cross in the $T-\mu_B$ plane, one can predict the location of the first-order line and the critical endpoint, as illustrated in the right panel of Fig.~\ref{fig:cartoon_firstorder}.

\begin{figure*}[t]
    \centering
    \includegraphics[width=\textwidth]{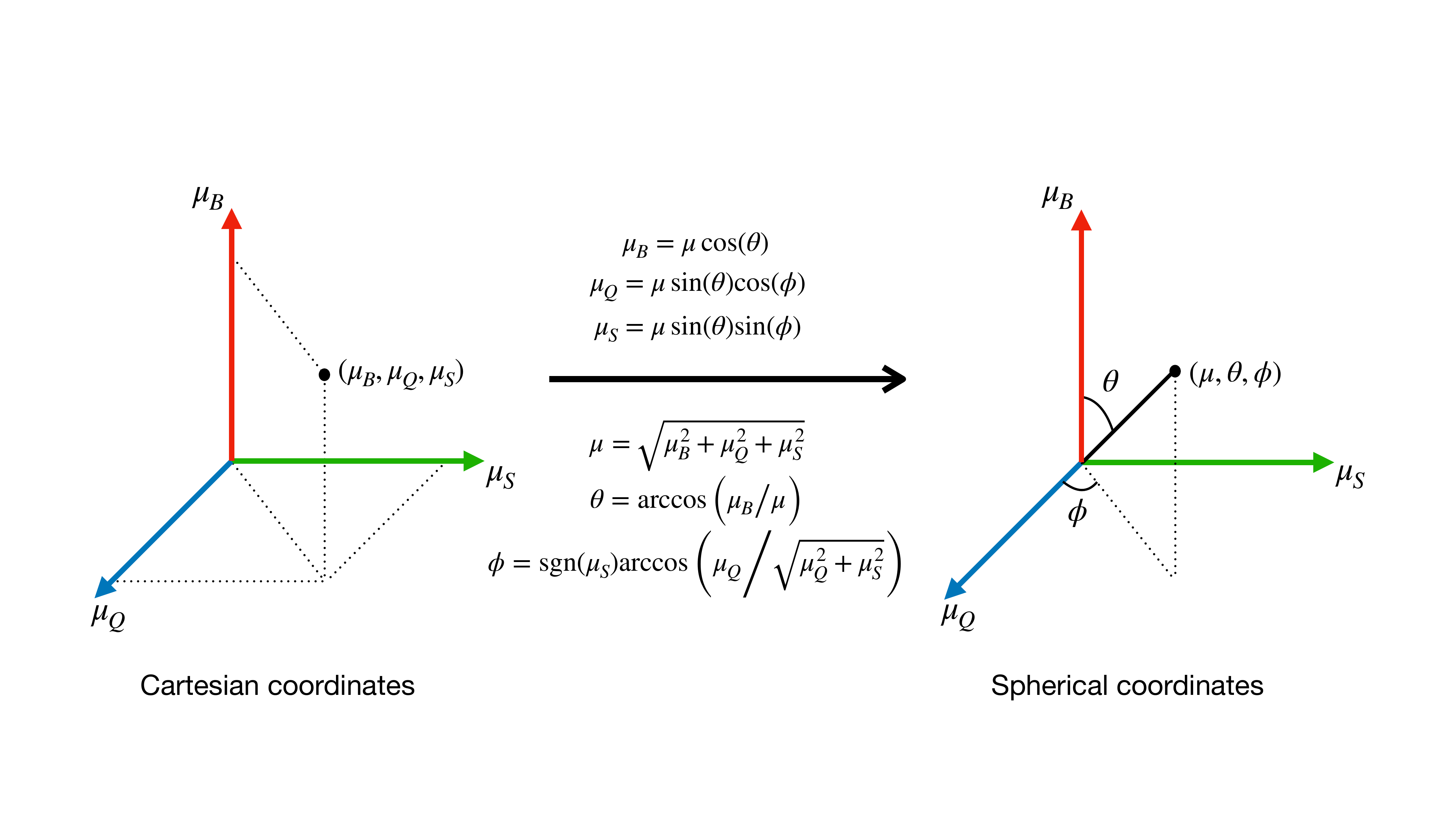}
    \vspace{-1.5cm}
    \caption{ The 3D space of chemical potentials mapped from cartesian coordinates $(\mus)$ (left) to spherical coordinates $(\mu,\theta,\varphi)$ (right). 
    }
    \label{fig:mapping}
\end{figure*}

Given that direct lattice QCD calculations are not feasible at finite $\mu_B$, a Taylor expansion of temperature $T$ around $\mu_B=0$, at constant $s$, is introduced to construct the entropy density contours:
\begin{equation}
    T_s(\mu_B;T_0)\approx T_0+\sum_{n=1}^N\alpha_{2n}(T_0)\frac{\mu_B^{2n}}{(2n)!}+\mathcal{O}(\mu_B^{2(N+1)}),
\end{equation}
where $T_0$ is the temperature at $\mu_B=0$, and the coefficients of expansion are written as
\begin{equation}
    \alpha_{2n}=\left(\frac{\partial^{2n}T}{\partial\mu_B^{2n}}\right)_s\Bigg\vert_{T=T_0,\mu_B=0}.
\end{equation}
Charge conjugation symmetry implies that $n_B=0$ at $\mu_B=0$, thus, all odd terms vanish. The explicit form for the second-order coefficient is
\begin{align}
    \alpha_2(T_0)
    &=\left(\frac{\partial^2 T}{\partial\mu_B^2}\right)_s\Bigg\vert_{T=T_0,\mu_B=0}
    \\
    &=-\frac{2T_0\chi_2^B(T_0)+T_0^2\chi_2^{B^\prime}(T_0)}{s^\prime(T_0)}.
\end{align}
Truncating the expansion at second-order yields:
\begin{equation}
\label{eq:TsO2}
    T_s(T_0;\mu_B)=T_0+\alpha_2(T_0)\frac{\mu_B^2}{2}.
\end{equation}

The inflection point of the entropy density at the CP can be described by the two conditions $(\partial T/\partial s)_{\mu_B}=0$ and $(\partial^2T/\partial s^2)_{\mu_B}=0$ which, in the specific case of this expansion, become $(\partial T_s/\partial T_0)_{\mu_B}=0$ and $(\partial^2T_s/\partial T_0^2)_{\mu_B}=0$. The first condition $(\partial T_s/\partial T_0)_{\mu_B}=0$ when applied on the expansion gives the critical baryon chemical potential $\mu_{B,c}$ as
\begin{equation}
    1+\alpha_2^\prime(T_{0,c})\frac{\mu_{B,c}^2}{2}=0
    \quad\Rightarrow\quad
    \mu_{B,c}=\sqrt{-\frac{2}{\alpha_2^\prime(T_{0,c})}}.
\end{equation}
The value of $T_{0,c}$ is identified by the second condition $(\partial^2T_s/\partial T_0^2)_{\mu_B}=0$:
\begin{equation}
    \alpha_2^{\prime\prime}(T_{0,c})=0.
\end{equation}
Both $\mu_{B,c}$ and $T_{0,c}$ can be inserted into Eq.~\eqref{eq:TsO2} to calculate the critical temperature $T_c$.

\subsection{Extension to four dimensions}

The full four-dimensional space of (2+1)-flavor QCD is spanned by the temperature $T$ and three conserved charge chemical potentials $(\mu_B,\mu_Q,\mu_S)$. 
Similarly to the approach employed in Ref.~\cite{Shah:2026qcs}, we use spherical coordinates in the chemical potential space as shown in Fig.~\ref{fig:mapping}. 
This mapping, first introduced in Ref.~\cite{Abuali:2025tbd}, reads
\begin{align}
    \mu_B&=\mu\cos\theta,
    \\
    \mu_Q&=\mu\sin\theta\cos\varphi,
    \\
    \mu_S&=\mu\sin\theta\sin\varphi.
\end{align}
In practice, we apply the entropy density contour method in the $T-\mu$ plane in a given direction in $\theta$ and $\varphi$, and this allows us to span the three-dimenional chemical potential space.

We define the generalized susceptibilities
\begin{equation}
    X_n^{\theta,\varphi}(T)=\frac{\partial^n(P(T,\mu,\theta,\varphi)/T^4)}{\partial(\mu/T)^n}\Bigg\vert_{\theta,\varphi},
\end{equation}
the first two being
\begin{align}
    X_1^{\theta,\varphi}(T)&=c_\theta\chi_1^B(T)+s_\theta c_\varphi\chi_1^Q(T)+s_\theta s_\varphi\chi_1^S(T),
    \\
    X_2^{\theta,\varphi}(T)&=
    c_{\theta}^2\chi_2^B(T)
    + s_{\theta}^2c_{\varphi}^2\chi_2^Q(T)
    + s_{\theta}^2 s_{\varphi}^2\chi_2^S(T) \nonumber \\
    & \quad + 2c_{\theta}s_{\theta}c_{\varphi}\chi_{11}^{BQ}(T)
    + 2c_{\theta}s_{\theta}s_{\varphi}\chi_{11}^{BS}(T)  \nonumber \\
    & \quad 
    + 2s_{\theta}^2c_{\varphi}s_{\varphi}\chi_{11}^{QS}(T)\, ,
    \label{eq:X's}
\end{align}
where $s_\gamma\equiv\sin\gamma$, $c_\gamma\equiv\cos\gamma$, and
\begin{equation}
    \chi_{ijk}^{BQS}=\frac{\partial^{i+j+k}P(T,\mus)/T^4}{\partial(\mu_B/T)^i\,\partial(\mu_Q/T)^j\,\partial(\mu_S/T)^k}.
\end{equation}

With these we construct the expansion coefficient generalized to four dimensions:
\begin{align}
    \alpha_2^{\theta,\varphi}(T_0)
    &=\left(\frac{\partial^2 T}{\partial\mu^2}\right)_{s,\theta,\varphi}\Bigg\vert_{T=T_0,\mu=0}
    \\
    &=-\frac{2T_0 X_2^{\theta,\varphi}(T_0)+T_0^2 X_2'^{\theta,\varphi}(T_0)}{s^\prime(T_0)},
\end{align}
which allows us to write the contour expansion truncated at second order as
\begin{equation}
\label{eq:TsO24D}
    T_s^{\theta,\varphi}(T_0,\mu)=T_0+\alpha_2^{\theta,\varphi}(T_0)\frac{\mu^2}{2}.
\end{equation}
With the four-dimensional generalization of the expansion established, the CP can be found by following the same procedure as in the two-dimensional case, with the direction in the four-dimensional space determined by $\theta$ and $\varphi$. 

\subsection{Lattice inputs}
To construct $X_2(T_0,\theta,\varphi)$, and thus the expansion coefficient $\alpha_2(T_0,\theta,\varphi)$, we use the parameterizations introduced in Ref.~\cite{Shah:2026qcs}.
For $\chi_2^B(T_0)$ and $s(T_0)$, we preserve the same parametrization as in Ref.~\cite{Shah:2024img}, which uses  the continuum extrapolated lattice results from Ref.~\cite{Borsanyi:2021sxv} and Ref.~\cite{Borsanyi:2013bia} respectively.  The rest of the conserved charge susceptibilities, in particular $\chi_2^Q(T_0)$, $\chi_{11}^{BQ}$, and $\chi_{11}^{BS}(T_0)$, are constructed to match the continuum-estimated second-order susceptibilities in Ref. \cite{Abuali:2025tbd}. To obtain $\chi_2^S(T_0)$ and $\chi_{11}^{QS}(T_0)$, we use:
\begin{align}
2\chi_{11}^{BQ}(T_0) - \chi_2^B(T_0) - \chi_{11}^{BS}(T_0) &= 0, \\
2\chi_{11}^{QS}(T_0) - \chi_2^S(T_0) - \chi_{11}^{BS}(T_0) &= 0.
\end{align}
These relations are enforced by isospin symmetry at $\mu=0$ which is imposed on the lattice for (2+1)-flavor QCD. These form the complete list of quantities required by $\alpha_2(T_0,\theta,\varphi)$ to perform the expansion of constant entropy density contours in the four-dimensional hypersphere. The expansion is naturally parameterized by $T_0$, the temperature at which each contour intersects the $\mu=0$ axis. To construct an equation of state, however, one requires thermodynamic quantities which are functions of the physical variables $(T,\mus)$. This is accomplished by numerically inverting Eq.~\eqref{eq:TsO24D} for the corresponding value of $T_0$ for every point in the four-dimensional space.

\section{Construction of the Four-Dimensional Equation of State}
\label{sec:eos}
In the previous section, we reviewed the entropy contour expansion in 2D, its extension to higher dimensions, and the input from lattice QCD at zero $\mu$ for lattice QCD susceptibilities.
In this section, we will use these susceptibilities and entropy density at vanishing $\mu$ to obtain the equation of state in four dimensions. 
Such a four-dimensional equation of state would be particularly useful in hydrodynamic simulations with local fluctuations of conserved charge densities~\cite{Schafer:2021csj,Pala:2025qoa,Monnai:2026fkp}.
Similar to the approach presented in Ref.~\cite{Shah:2026pue}, we first calculate pressure from entropy density at finite temperature and chemical potential, and then calculate the other thermodynamics. 

\subsection{Pressure from entropy density}

The contours of constant entropy density can be constructed using the four-dimensional expansion~\eqref{eq:TsO24D} for a set of $T_0$ values, anchored at $\mu=0$, and spanning the chemical potential space in $\theta$ and $\varphi$. By inverting Eq.~\eqref{eq:TsO24D} to obtain $T_0^{\theta,\varphi}(T,\mu)$, we compute the value of the entropy density $s^{\theta,\varphi}(T,\mu)$ at every direction in the four-dimensional space. This function is then converted from spherical to cartesian coordinates to obtain $s(T,\boldsymbol{\mu})$, where $\boldsymbol{\mu}=(\mus)$.

Given that $s=[\partial P(T,\boldsymbol{\mu})/\partial T]_{\boldsymbol{\mu}}$, we obtain pressure by integrating the entropy density with respect to temperature while holding chemical potentials fixed:
\begin{align}
\label{eq:pressure}
    \int_{T\low}^T d\tilde T s(\tilde T,\boldsymbol{\mu})=P(T,\boldsymbol{\mu})-P(T\low,\boldsymbol{\mu}),
\end{align}
where $\tilde T$ is the integration constant. Here we set $T\low=80$ MeV and calculate $P(T\low,\boldsymbol{\mu})$ using the Quantum van der Waals Hadron Resonance Gas (QvdW-HRG) model~\cite{Vovchenko:2016rkn}, supplemented with an effective mass model for pion and kaon condensation~\cite{Vovchenko:2020crk}. 
The QvdW-HRG model provides a description of low-temperature and high chemical potential hadronic matter, which also reproduces lattice-QCD thermodynamics at $\mu=0$. 
This model is chosen because the expansion itself does not accurately describe the dilute hadronic phase, where the entropy-density contours deviate from the quadratic approximation~\cite{Marczenko:2025znt,Shah:2026pue}. 
Hence, using this model as an integration constant at low $T$ 
helps in describing the physics in the high chemical potential and low-temperature regime.

\subsection{Conserved charge densities}
Utilizing the pressure at finite chemical potentials, the three conserved charge densities $n_B$, $n_Q$, $n_S$ are computed by differentiating it with respect to the corresponding chemical potential: $n_i=\left(\partial P/\partial\mu_i\right)_{T,\mu_j,\mu_k}$, with $i=B,Q,S$ and $i\neq j\neq k$. The densities read
\begin{align}
    & n_B(T,\boldsymbol{\mu})
    =n_{B_\text{HRG}}(T_\text{low},\boldsymbol{\mu}) \nonumber
    \\
    & \quad +\bigg[\mu_B \tilde T_0^2 X_2 
    -\frac{\sqrt{\mu_Q^2+\mu_S^2}}{2} \tilde T_0^2\dth{X_2}\bigg]\boundary,
    \label{eq:nB}
\end{align}
\begin{align}
    n_Q(T,\boldsymbol{\mu})
    =&n_{Q_\text{HRG}}(T_\text{low},\boldsymbol{\mu}) \nonumber
    \\
    &+\bigg[\mu_Q \tilde T_0^2 X_2+\frac{\mu_B\mu_Q}{2\sqrt{\mu_Q^2+\mu_S^2}} \tilde T_0^2\dth{X_2} \nonumber
    \\
    &-\frac{\mu_S\mu^2}{2\sqrt{\mu_Q^2+\mu_S^2}^2} \tilde T_0^2\dph{ X_2}\bigg]\boundary,
    \label{eq:nQ}
\end{align}
\begin{align}
\label{eq:nS}
    n_S(T,\boldsymbol{\mu})
    =&n_{S_\text{HRG}}(T_\text{low},\boldsymbol{\mu}) \nonumber
    \\
    &+\bigg[\mu_S \tilde T_0^2 X_2+\frac{\mu_B\mu_S}{2\sqrt{\mu_Q^2+\mu_S^2}}\tilde T_0^2 \dth{ X_2} \nonumber
    \\
    &+\frac{\mu_Q\mu^2}{2\sqrt{\mu_Q^2+\mu_S^2}^2}\tilde T_0^2 \dph{ X_2}\bigg]\boundary,
\end{align}
where $X_2$ is defined in Eq.~\eqref{eq:X's}. The full derivation of the conserved charge densities can be found in the Appendix~\ref{app:derivations}. With the three conserved charge densities, the energy density can be calculated from
\begin{equation}
   \epsilon(T,\boldsymbol{\mu})=-P+Ts+\sum_i\mu_in_i, \qquad i\in\{B,Q,S\}.
\end{equation}

\subsection{Higher order derivatives of the pressure}
Utilizing the charged densities, we can now calculate the second order susceptibilities at finite chemical potentials. The three diagonal and three off-diagonal
second-order susceptibilities are collected in the symmetric matrix:

\begin{equation}
    T^2 \chi_{ij}(T,\boldsymbol{\mu})
    =\left(\frac{\partial n_i}{\partial\mu_j}\right)_{T,\mu_{k\neq j}}, \quad
    i,j\in\{B,Q,S\},
    \label{eq:chis}
\end{equation}
whose diagonal entries are the diagonal susceptibilities
$\chi_2^i\equiv\chi_{ii}$, with $i=j$, and whose off-diagonal entries are the correlators $\chi_{11}^{ij}$, with $i\neq j$.
The higher-order diagonal susceptibilities 
along a single charge direction are obtained by:
\begin{equation}
\label{eq:chin_general}
    T^{4-n}\chi_n^i(T,\boldsymbol{\mu})
    =\left(\frac{\partial^{\,n-2}(T^2\chi_2^i)}{\partial\mu_i^{\,n-2}}\right)_{T,\mu_{j\neq i}},~~
    i\in\{B,Q,S\},
\end{equation}
for $n\geq2$.
In this work, we evaluate the third- and fourth-order susceptibilities
$\chi_{3}^{i}$ and $\chi_{4}^{i}$ for $i=B,Q,S$.
With the charged susceptibilities obtained using Eq. (\ref{eq:chis}), the three charge densities, Eqs. (\ref{eq:nB}-\ref{eq:nS}), and the entropy density, we can construct the full Hessian
\begin{align}
    \hat{H}=\frac{\partial^2 P}{\partial(T,\mu_B,\mu_Q,\mu_S)^2}=
    \begin{pmatrix}
        s_T & \mathbf{b}^T \\
        \mathbf{b} & \hat{A}
    \end{pmatrix},
\end{align}
where $s_T=(\partial s/\partial T)_{\mu}$, $\mathbf{b}\equiv(\partial\mathbf{n}/\partial T)_{\mu_B,\mu_Q,\mu_S}$, and $\hat{A}=T^2\chi$.

To calculate the heat capacity we begin with the standard thermodynamic relation,
\begin{equation}
    c_V(T,\boldsymbol{\mu})=T\left(\frac{\partial s}{\partial T}\right)_{\mathbf{n}},
\end{equation}
where $\mathbf{n}=(n_B,n_Q,n_S)$, and which we can express in terms of the Hessian~(see Appendix~\ref{app:hessian}):
\begin{align}
    c_V(T,\boldsymbol{\mu})=T\frac{\det\hat{H}}{\det\hat{A}}.
\end{align}

To calculate the speed of sound squared, we use the following form in terms of the inverse Hessian matrix~(see Appendix~\ref{app:hessian})
\begin{align}
    c_s^2(T,\boldsymbol{\mu})=\left(\frac{\partial P}{\partial\epsilon}\right)_{\mathbf{n}/s} =\frac{\sum_{i,j=0}^2x_i\hat{H}_{ij}^{-1}x_j}{(\epsilon+P)},
\label{eq:cs2_final}
\end{align}
where $\mathbf{x}=(T,n_B,n_Q,n_S)$.

\begin{figure*}[t]
    \centering
    \includegraphics[width=\textwidth]
        {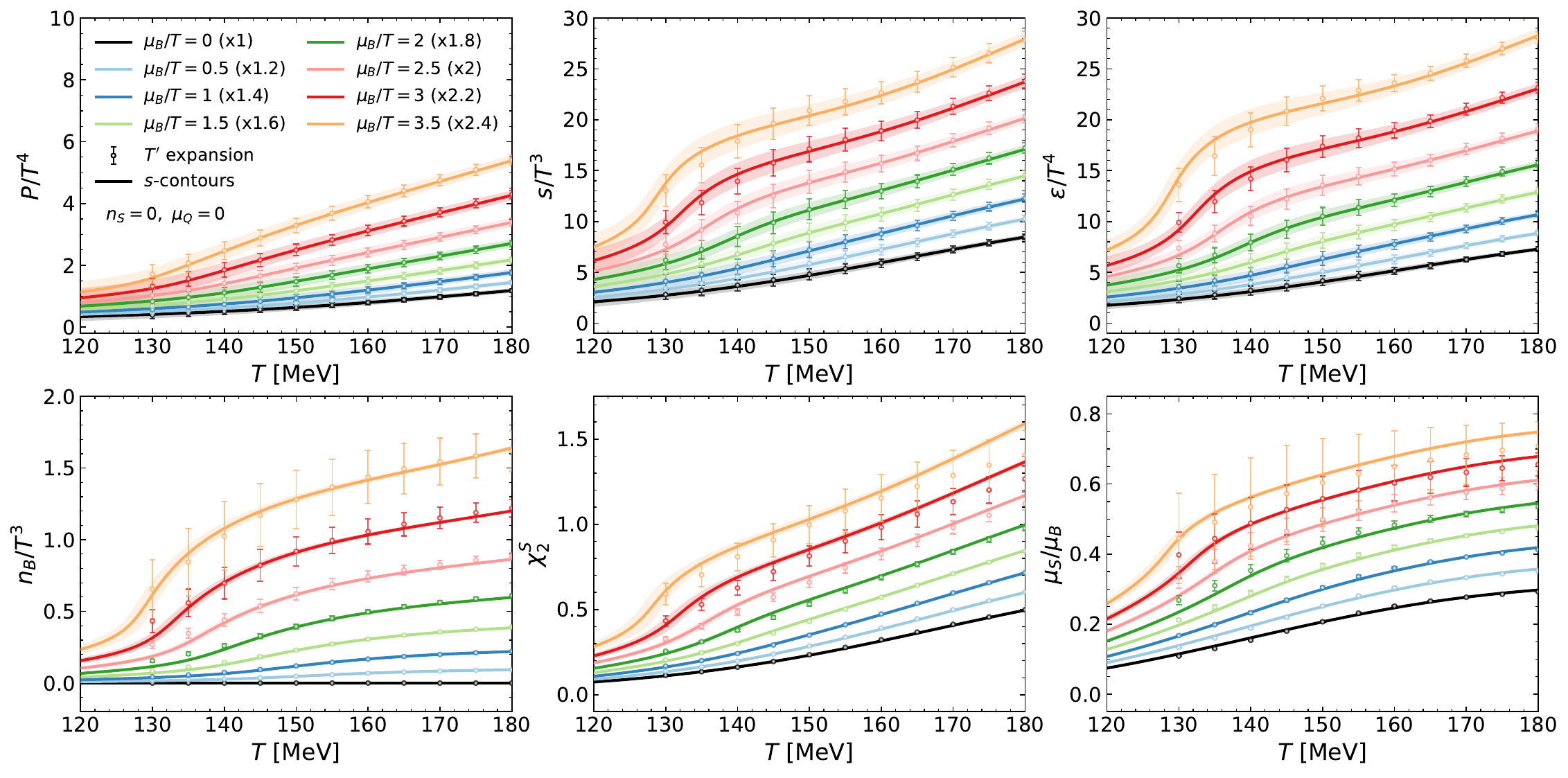}
    \caption{Normalized pressure (top left), entropy density (top middle), energy density (top right), baryon density (bottom left), second order strangeness susceptibility (bottom middle) and strangeness chemical potential-to-baryon chemical potential ratio (bottom right) shown as a function of temperature for different slices of $\mu_B/T$ with strangeness-neutral conditions $(n_S=0,\mu_Q=0)$. Solid bands represent entropy contours expansion scheme which is compared to the T'-expansion scheme from Ref.~\cite{Borsanyi:2022qlh} displayed by the error bars. All the quantities are scaled by a factor for each $\mu_B/T$ slice to avoid overlapping error bars. We observe agreement within the two expansion schemes up to $\mu_B/T \leq 3.5$.
    }
    \label{fig:sn_compare}
\end{figure*}

\begin{figure*}
    \centering
    \includegraphics[width=\textwidth]
        {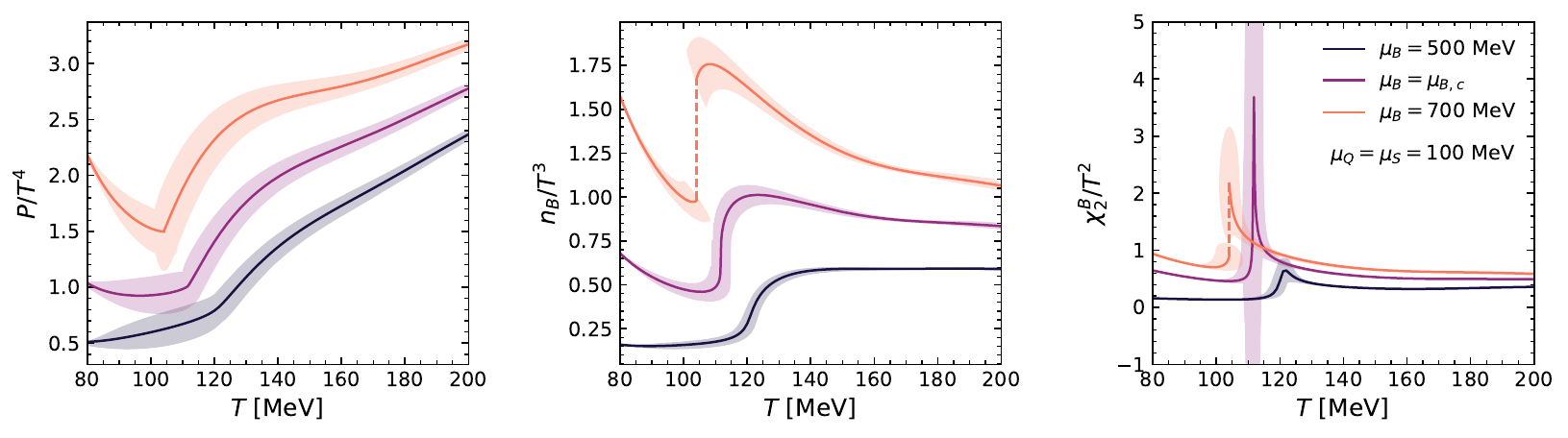}
    \caption{Scaled pressure (left), baryon density (middle), and the second-order baryon susceptibility (right) as functions of the temperature at various values of $\mu_B$ with fixed $\mu_Q=\mu_S=100$ MeV. The black curve shows $\mu_B$ corresponding to the crossover region, magenta bands illustrate $\mu_B$ at the CP, and orange bands display the baryon chemical potential in the first-order phase transition region.}
    \label{fig:gen_fig_2}
\end{figure*}

\section{Results}
\label{sec:results}

\begin{figure*}[t]
    \centering
    \includegraphics[width=\textwidth]
        {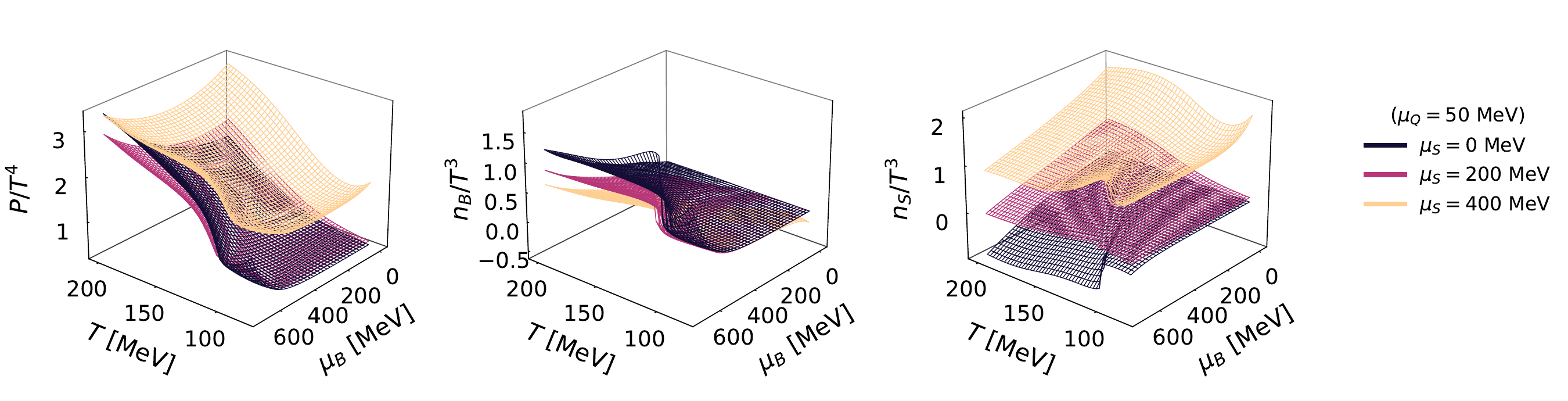}
    \caption{From left to right, the
    panels show the normalized pressure, baryon density  and strangeness density as functions of T and $\mu_B$. The electric charge chemical potential $\mu_Q$ is fixed at 50 MeV. The colors represent different strangeness chemical potential where $\mu_S=0$ MeV is displayed in black, $\mu_S=200$ MeV in magenta and the orange is at $\mu_S=400$ MeV. The figure illustrates that one can construct a four dimensional equation of state with expansion of contours of constant entropy density.   }
    \label{fig:eos_muS_layers}
\end{figure*}
In this section, we present the resulting four-dimensional equation of state constructed from constant entropy density contours. 
The equation of state depends on the full set of conserved-charge chemical potentials $(\mus)$ and contains a critical surface separating crossover and first-order regions of the phase diagram. 
We first discuss its general properties before considering applications to physically relevant trajectories corresponding to heavy-ion collisions, the early Universe, and astrophysical environments. The uncertainties of the equation of state are calculated by propagating the uncertainties of the lattice QCD parameterization through the contour construction. This is accomplished via linear error propagation using forward-mode automatic differentiation, from which confidence intervals are determined for all thermodynamic observables. 

Before reporting on the results of the thermodynamics, we mention the caveats of the expansion.
First, since we expand entropy density contours, by construction our method cannot correctly describe Bose--Einstein condensation.
Our approach captures phase transitions through $s = (\partial P/\partial T)_{\mu}$, a first derivative of the pressure: a first-order transition appears as a discontinuity in $s$, arising when the entropy contours become multivalued, with the CP marking its onset.
Bose--Einstein condensation, by contrast, is a continuous transition in which $s$ remains continuous and the non-analyticity appears only in higher derivatives.
In QCD, pion condensation is expected to occur at $\mu_Q \gtrsim m_\pi$ and $T \lesssim 160$~MeV~\cite{Brandt:2017oyy}. 
Similarly, kaon condensation may take place at large $\mu_S \sim m_K$~\cite{Kaplan:1986yq,Kaplan:1987sc,Schafer:2001bq}.
Our entropy contour expansion does not incorporate the singularity associated with a possible formation of a condensate.
However, we do use the QvdW-HRG model to fix the pressure integration constant at $T = 80$~MeV, which incorporates an effective mass model for pions and kaons that provides a thermodynamically consistent description of the Bose-Einstein condensation as a second-order phase transition~\cite{Savchuk:2020yxc,Vovchenko:2020crk}.
Thus, our construction effectively contains a singularity in the chemical potential directions associated with pion and kaon condensation through the integration constant.
This does not, however, mean that condensation is described by the equation of state consistently. 

Another caveat of the equation of state is the fact that the expansion is truncated at $\mathcal{O}(\mu_B^2)$, and we do not take into account the 
systematic error due to the truncation of the expansion in this work.
Going to the next order becomes challenging, requiring higher-order conserved charge susceptibilities (up to fourth order) and their temperature derivatives, which are not yet available from the lattice simulations with sufficient precision. Including higher-order terms in the expansion in the future may modify the results depending on its convergence properties, which is beyond the scope of this work.
Consequently, the present results describe a scenario where the CP in the pure $\mu_B$ direction occurs at $\mu_B \sim 600$~MeV, as predicted by functional methods and holography, but they do not constitute a definitive prediction of the CP existence and location.

\subsection{Handling the first-order phase transition region}

The present expansion permits multi-valued behavior of various thermodynamic quantities at fixed temperature and chemical potentials,
as was observed previously in the two-dimensional equation of state constructed in Ref. \cite{Shah:2026pue}. 
The multi-valuedness is due to the fact that we use an analytic expansion truncated at second-order and it yields a mean-field description of thermodynamics for a first-order phase transition, which contains both the metastable and unstable branches of the solutions.

The physical equation of state can be obtained by implementing a mixed-phase construction. The exact implementation depends on external conditions, namely on whether chemical potentials and densities are held fixed. Without either construction performed, the EoS from the expansion naturally provides thermodynamics with the stable, metastable, and unstable branches.

\subsubsection{Maxwell construction at fixed $\theta$ and $\phi$}

When working at fixed angles $\theta$ and $\phi$, the values of all three chemical potentials are fixed at any point in the $T$-$\mu_B$ phase diagram.
If multiple solutions exist at a single point, the equilibrium phase is given by the one with the maximum pressure, in accordance with the Gibbs criterion.
Phase coexistence is given by the equality of pressures of the two phases,
\begin{align}
P_{\rm (1)}^{\theta,\phi}(T,\mu_B) = P_{\rm (2)}^{\theta,\phi}(T,\mu_B),
\end{align}
which defines a line on the $T$-$\mu_B$ phase diagram for fixed values of $\theta$ and $\phi$~(or, alternatively, for fixed values of $\mu_Q$ and $\mu_S$).
In our description, $P_{\rm (1)}^{\theta,\phi}(T,\mu_B)$ and $P_{\rm (2)}^{\theta,\phi}(T,\mu_B)$ are described by the same multi-valued function $P^{\theta,\phi}(T,\mu_B)$ obtained implicitly from the constant entropy contour expansion.
The practical algorithm for finding the coexistence line, as well as stable, metastable, and unstable branches, is given by the Maxwell construction of equal areas applied to entropy density $s^{\theta,\phi}(T;\mu_B)$ at a constant value of $\mu_B$.
Here $T = T_s^{\theta,\phi}(T_0;\mu)$ is parametrized by $T_0$. 
The algorithm is identical to the pure $\mu_B$ case and described in Ref.~\cite{Shah:2024img}.

\subsubsection{Gibbs construction at fixed $n_S$ and $n_Q/n_B$}

The physical conditions corresponding to different systems, for example in heavy-ion collisions, are imposed on conserved charge densities rather than on chemical potentials.
For instance, heavy-ion collisions are characterized by fixed global ratios of charge-to-baryon number ratio, $n_Q / n_B \simeq 0.4$ and strangeness neutrality, $n_S = 0$.
Inside the mixed phase, equilibrium conditions require the two phases to have equal pressures, temperatures, and chemical potentials.
However, the two phases need not have fixed ratios of charge-to-baryon density, or vanishing strangeness~~\cite{Greiner:1987tg,Glendenning:1992vb,Hempel:2013tfa,Poberezhnyuk:2018mwt}. 
Instead, in the Gibbs construction, the density constraints are applied only on the total densities from the two phases.

Consider now phase coexistence where the dilute phase~(hadron gas) occupies volume fraction $\lambda$ of the full volume.
The total densities are then given by $n_{B,Q,S}^{\rm mix}(T,\boldsymbol{\mu};\lambda) = (1-\lambda) n_{B,Q,S}^{(1)}(T,\boldsymbol{\mu}) + \lambda n_{B,Q,S}^{(2)}(T,\boldsymbol{\mu})$.
The mixed phase conditions read
\begin{align}
\label{eq:ccm1}
P_{(1)}(T,\boldsymbol{\mu}) & = P_{(2)}(T,\boldsymbol{\mu}), \\
\label{eq:ccm2}
\frac{n_Q^{\rm mix}(T,\boldsymbol{\mu};\lambda)}{n_B^{\rm mix}(T,\boldsymbol{\mu};\lambda)} & = \frac{n_Q}{n_B}, \\
\label{eq:ccm3}
n_S^{\rm mix}(T,\boldsymbol{\mu};\lambda) & = 0.
\end{align}
Because volume fraction can be varied in a region $0 \leq \lambda \leq 1$, one obtains a mixed-phase region in the $T$-$\mu_B$ plane as opposed to a line in the case of fixed chemical potentials.
The boundaries of the mixed phase region can be determined by solving Eqs.~\eqref{eq:ccm1}-\eqref{eq:ccm3} for $\lambda = 0$ and $\lambda = 1$.
These boundaries are shown in our calculations below on the $T$-$\mu_B$ diagram.

\subsection{General properties}
\label{sec:general_results}

Using the thermodynamic calculations from Section \ref{sec:eos}, we obtain the equation of state for multiple conserved charges of QCD. 
We first compare our results to available lattice QCD benchmarks at small chemical potentials.
In the pure-$\mu_B$ direction, comparisons with the $T'-$expansion scheme~\cite{Borsanyi:2021sxv} were presented in Ref.~\cite{Shah:2026pue}, indicating good agreement throughout $\mu_B/T < 3.5$.
Here we perform comparisons under SN conditions, with the corresponding lattice results within the $T'-$expansion from Ref.~\cite{Borsanyi:2022qlh}. 
To enforce strangeness neutrality, we solve for the conditions in our expansion where $n_S=0$, while keeping $\mu_Q=0$ in line with the setup in Ref.~\cite{Borsanyi:2022qlh}. 

Figure~\ref{fig:sn_compare} presents the comparison, where we plot pressure, entropy density, energy density, baryon density, strangeness susceptibility, and the $\mu_S/\mu_B$ ratio as functions of the temperature for different values of $\mu_B/T$. 
The empty circles show the lattice results using the $T'-$ expansion scheme with the associated errors, while the EoS from the expansion of entropy contours  is shown by the solid lines with the surrounding bands showing propagated errors.
The thermodynamic quantities at different $\mu_B/T$ ratios are scaled to particular values, shown in brackets in the legend, for better visibility. We observe good agreement across all the thermodynamic quantities among both the expansion schemes. We note that the uncertainties for densities and susceptibilities are significantly smaller for the entropy contour expansion than for the $T'$-expansion scheme. The reason lies in the truncation of the expansions at second order: in the $T'$-expansion scheme, the second-order coefficient involves fourth-order susceptibilities, which carry larger error bars. In contrast, $\alpha_2$ depends only on the second-order susceptibility and the entropy density, both of which are known more precisely. Moreover, in the analytical expressions for densities and susceptibilities at finite chemical potential, the explicit contribution of $s(T_0)$ cancels (see Secs.~\ref{sec:densitiy_derivs} and \ref{sec:sus_derivs} of the supplemental material). The entropy density still enters implicitly, however, since the expansion point $T_0$ corresponding to a given $(T,\mu_B)$ is obtained by inverting $T = T_0 + \alpha_2(T_0)\,\mu^2/2$, so that the uncertainty of $\alpha_2$, including that of $s$, propagates into $T_0$.

Figure~\ref{fig:gen_fig_2} displays the normalized pressure, baryon density, and the second-order baryon susceptibility as functions of $T$ at various values of $\mu_B$ and fixed non-zero  $\mu_Q = \mu_S = 100$~MeV.
The three different values of $\mu_B$ are chosen to represent the crossover region, the vicinity of the CP, and a first-order phase transition region, respectively.
We observe that the susceptibilities diverge as we move towards the CP. In the present case the CP in $(T,\mu_B)$ is at $(112\pm8,616\pm72)$~MeV.

Figure~\ref{fig:eos_muS_layers} displays the 3D structure of the equation of state as a function of $(T,\mu_B)$ for different values of $\mu_S$ at a fixed $\mu_Q$ value. In particular, we observe the behavior of 
normalized pressure, baryon density and strangeness density for three different values of $\mu_S=0$, $200$, and $400$~MeV, shown by the black, magenta and orange curves. One can observe that, when $\mu_S =0$, the strangeness density $n_S$ is actually negative for increasing values of $\mu_B$, but as $\mu_S$ increases, it smoothly grows larger and becomes positive.

The present equation of state can be used to analyze any direction in the four-dimensional phase diagram of QCD, and is especially useful for hydrodynamic simulations, where conserved charge densities can be simultaneously finite and vary locally.
We now shift our focus to a number of physically relevant cases where the 4D EoS can be applied. 

\subsection{Heavy-ion collision conditions}
\label{sec:hic_results}

\begin{figure}[t]
    \includegraphics[width=\columnwidth]
        {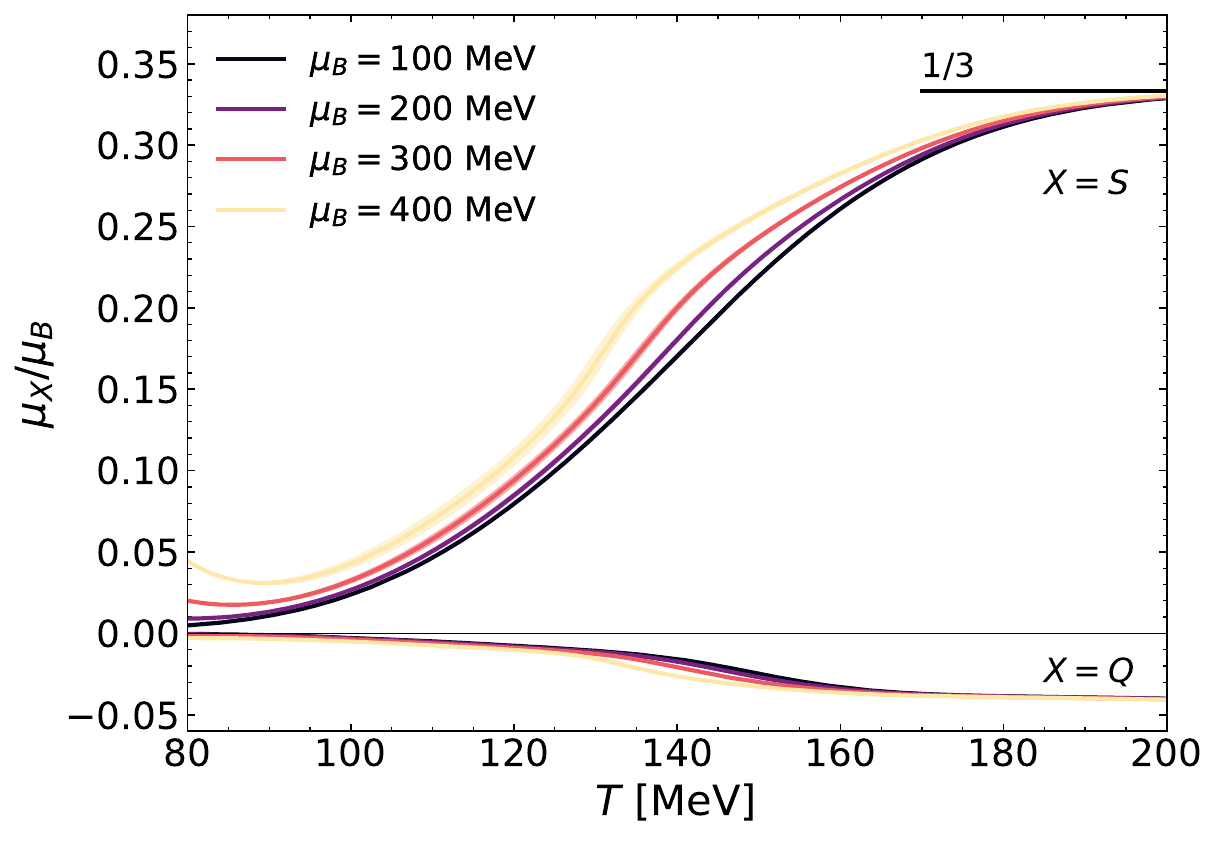}
    \caption{Chemical potential ratios, $\mu_X/\mu_B$, displayed as functions of the temperature for different values of $\mu_B$ under heavy-ion collision constraints ($n_S=0$ and $n_Q/n_B=0.4$). In the positive vertical axis, $X=S$ i.e. we show the ratio  $\mu_S/\mu_B$, while in the negative vertical axis $X=Q$ i.e. we show the ratio  $\mu_Q/\mu_B$.
    }
    \label{fig:muSovermuB}
\end{figure}

\begin{figure}[b]
    \includegraphics[width=\columnwidth]
        {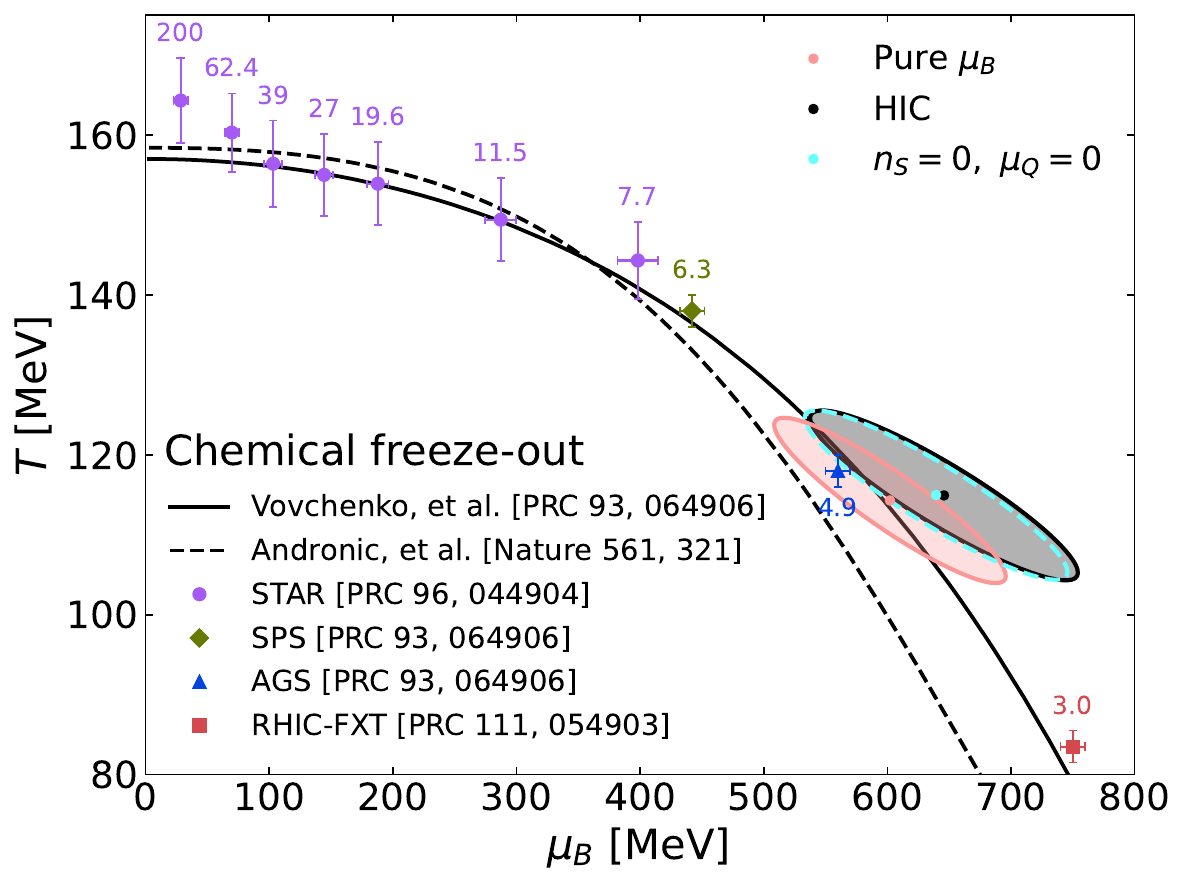}
    \caption{The solid black ellipse indicates the 68\% confidence interval region for the CP under heavy-ion collision constraints, which is shifted to higher $\mu_B$ compared to the prediction at $\mu_Q= \mu_S=0$ shown by the magenta ellipse. The light blue ellipse shows the 68\% confidence region for the CP in the simplified strangeness-neutral case $n_S=0$, $\mu_Q=0$. The solid and  dashed black curves are freeze-out lines  for heavy-ion collisions from Ref.~\cite{Andronic:2017pug} and Ref.~\cite{Vovchenko:2015idt}, respectively.
    The differently colored symbols represent individual chemical freeze-out points for different heavy-ion experiments~\cite{STAR:2017sal,Lysenko:2024hqp}
    }
    \label{fig:HIC_FOPT}
\end{figure}

\begin{figure*}[t]
    \centering
    \includegraphics[width=\textwidth]
        {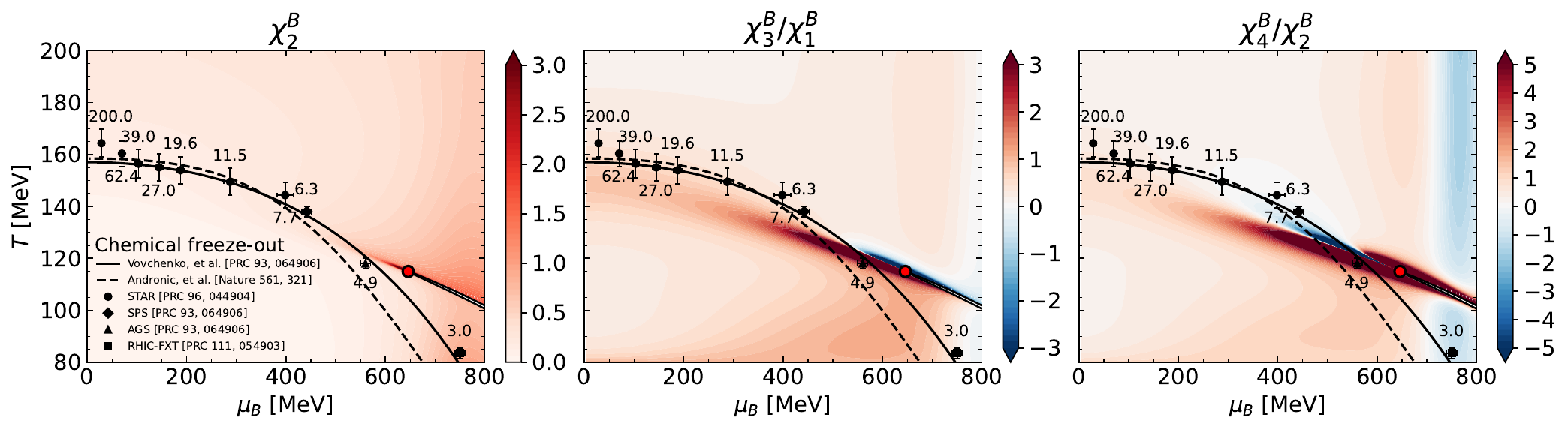}
    \caption{Contour maps of constant second-order baryon susceptibility (left), ratio of third to first baryon susceptibility (middle), and ratio of fourth to second baryon susceptibility (right) shown on the $T-\mu_B$ plane for heavy-ion conditions. The solid and dashed black curves are heavy-ion chemical freeze-out lines from Ref.~\cite{Andronic:2017pug} and Ref.~\cite{Vovchenko:2015idt}, respectively.  
    The differently shaped black points represent chemical freeze-out conditions for different heavy-ion experiments~\cite{STAR:2017sal,Lysenko:2024hqp}.
    }
    \label{fig:hic_contours}
\end{figure*}

\begin{figure*}
    \centering
    \includegraphics[width=\textwidth]
        {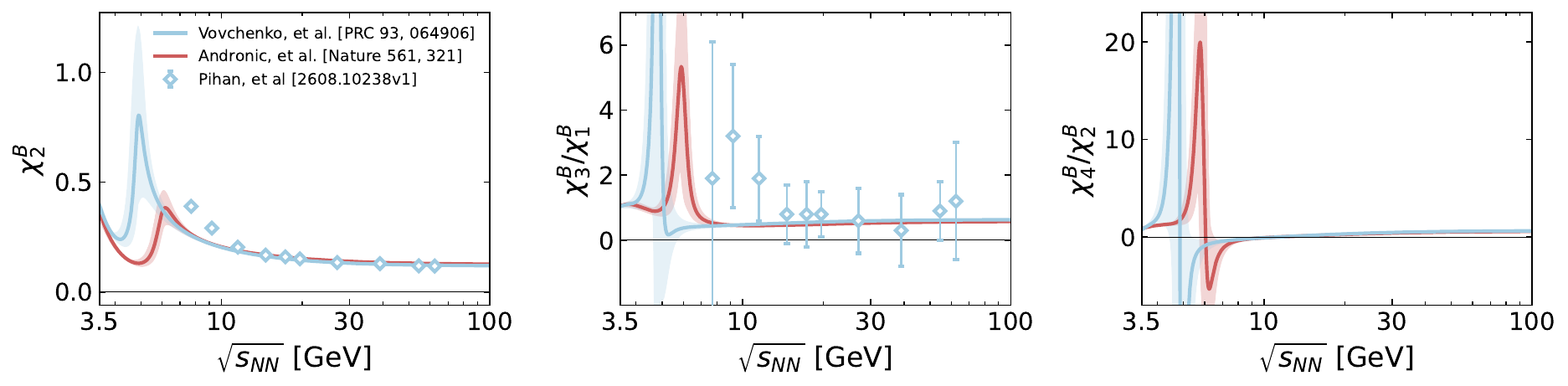}
    \caption{ Second order baryon susceptibility (left), ratio of third to first baryon susceptibility (middle) and ratio of fourth to second baryon susceptibility (right) calculated on two different freeze-out lines as functions of the collision energy. The blue and red bands display the quantity calculated on the freeze-out line from Refs.~\cite{Vovchenko:2015idt} and Ref.~\cite{Andronic:2017pug}, respectively. The blue dots with error bars are the extracted points for
    $\chi_2^B$ and $\chi_3^B/\chi_1^B$
    from net proton measurements computed in Ref.~\cite{Pihan:2026xng} based on hydrodynamics simulations.
    }
    \label{fig:freezeout_ratios}
\end{figure*}

Heavy-ion collisions are typically characterized by strangeness neutrality, i.e. $n_S = 0$, in general  requiring $\mu_S\neq 0$, as well as by a fixed charge fraction, i.e. the ratio between the electric charge and baryon densities, which depends the colliding nuclei. For Pb-Pb collisions, we therefore impose the conditions below to define the relevant slice of the phase diagram:
\begin{equation}
    n_S=0,
    \qquad
    n_Q=0.4\,n_B.
    \label{eq:hic_conditions}
\end{equation}
Hence at each point in $(T,\mu_B)$, we find the corresponding $\mu_Q$ and $\mu_S$ values that together satisfy the conditions in Eq.~\eqref{eq:hic_conditions}. The four-dimensional equation of state is thereby reduced to an effective two-dimensional surface parameterized by $T$ and $\mu_B$ for fixed densities. We will now focus on various quantities calculated by imposing these heavy ion conditions.

\subsubsection{Chemical potential ratios}
Figure~\ref{fig:muSovermuB} displays the behavior of chemical potential ratios $\mu_S/\mu_B$ and $\mu_Q/\mu_B$ as functions of temperature for different values of $\mu_B$. 
At low temperatures, strangeness neutrality is maintained with relatively low $\mu_S$ due to strange hadrons being thermally suppressed. 
As the temperature passes through the crossover region, the ratio $\mu_S/\mu_B$ increases rapidly reflecting an increase in the contribution of strange degrees of freedom. 
A similar $\mu_S/\mu_B$ ratio behavior is also seen in other expansions from lattice QCD~\cite{Borsanyi:2022qlh,Bollweg:2024epj}. 
At high temperatures, all trajectories approach approximately the Stefan-Boltzmann limit $\mu_S/\mu_B=1/3$ where strangeness neutrality corresponds to setting the strange-quark chemical potential to be zero. 
The ratio $\mu_Q/\mu_B$ shows small negative values, reflecting the relatively small isospin asymmetry induced by the colliding nuclei.

\subsubsection{Heavy-ion critical point}
Under HIC conditions, Eq.~\eqref{eq:hic_conditions}, we find the location of the CP to be at $(T_c,\mu_{B,c})=(115\pm7, 646\pm72)$ MeV, with the critical charge and strangeness chemical potentials corresponding to $(\mu_{Q,c},\mu_{S,c})=(-12\pm3,116\pm16)$.
Taking into account correlated input uncertainties,
this shifts the CP by $(\Delta T_c,\Delta\mu_{B,c})=(0.6\pm0.6,44\pm11)$ MeV as compared to the 
pure-$\mu_B$ CP, where $(T_c,\mu_{B,c})=(114 \pm 7, 602 \pm 62)$ MeV~\cite{Shah:2024img}, leaving the critical temperature essentially unchanged.
Fig.~\ref{fig:HIC_FOPT} compares these CP estimates, including their corresponding 68\% confidence interval ellipses, on the $T$-$\mu_B$ phase diagram.
We also examine the CP location relative to the SN case, at $(T,\mu_B)=(115\pm7,639\pm70),$ with a $(\Delta T_c,\Delta\mu_{B,c})=(-0.04\pm0.03,6\pm1)$ shift from the heavy-ion collision CP. We note that these CPs approximately overlap as a result of the small contribution of $\mu_Q$ after heavy-ion collisions are imposed. These critical points, as well as the charge-neutral (CN) critical point $(n_Q=0,~\mu_S=0)$ are written in Table~\ref{tab:CP_locations}.
We also show chemical freeze-out points at different energies from the literature~\cite{Vovchenko:2015idt,STAR:2017sal,Lysenko:2024hqp} along with two chemical freeze-out parameterizations~\cite{Vovchenko:2015idt,Andronic:2017pug}.
The differences between the two lines provide an estimate of the uncertainty in the freeze-out line, especially toward smaller collision energies.
In the conventional picture, the chemical freeze-out line describes a chemically equilibrated hadron resonance gas at the late stage of the collisions, and thus serves as a lower bound on the temperature for the possible location of the CP~\cite{Lysenko:2024hqp}.
We observe that both freeze-out lines pass well below the CP location under HIC and strangeness-neutral conditions
Ref.~\cite{Vovchenko:2015idt}.
The chemical freeze-out at collision energies $\sNN \sim 3-5$~GeV appears to be in the closest vicinity to the CP.

\begin{table*}
\label{tab:critical_points}
\centering
\begin{tabular}{|c|c|c|c|c|c|}
\hline
 & $T_{0c}$ (MeV) & $T_c$ (MeV) & $\mu_{B,c}$ (MeV) & $\Delta T_c$ (MeV) & $\Delta\mu_{B,c}$ (MeV) \\
\hline
Pure $\mu_B$ & $140.9 \pm 1.9$ & $114.3 \pm 6.9$ & $602.1 \pm 62.1$ & $0.0 \pm 0.0$ & $0.0 \pm 0.0$ \\
HIC & $141.0 \pm 1.9$ & $114.9 \pm 7.0$ & $645.8 \pm 71.8$ & $0.6 \pm 0.6$ & $43.7 \pm 11.1$ \\
SN & $141.0 \pm 1.9$ & $115.0 \pm 7.0$ & $639.4 \pm 70.4$ & $0.7 \pm 0.6$ & $37.3 \pm 9.8$ \\
CN & $140.9 \pm 1.9$ & $113.7 \pm 7.1$ & $612.3 \pm 65.0$ & $-0.5 \pm 0.3$ & $10.2 \pm 3.3$ \\
\hline
\end{tabular}
\caption{The location of the critical point for different physical conditions which includes pure $\mu_B$, heavy ion collision (HIC), $\mu_Q=0$ with strangeness neutral (SN) and $\mu_S=0$ with charge neutral (CN) conditions .}
\label{tab:CP_locations}
\end{table*}

\subsubsection{Baryon number susceptibilities}
Equilibrium fluctuations of conserved charges, in particular baryon number, are expected to show distinctive features in the vicinity of the CP ~\cite{Stephanov:2008qz,Stephanov:2011pb,Karthein:2025hvl}. 
Figure~\ref{fig:hic_contours} shows the contour plots of $\chi_2^B$, $\chi_3^B/\chi_1^B$, and $\chi_4^B/\chi_2^B$ for the mean value of the parametrization calculated under HIC conditions in the $T-\mu_B$ plane, along with the freeze-out estimates discussed in the previous subsection.
Figure~\ref{fig:hic_contours} also shows the first-order coexistence region, obtained via the Gibbs construction.
In contrast to the $\mu_B$-only case, where the first-order region collapses into a line on the $T$-$\mu_B$ plane, here the binodals corresponding to the dilute and dense phases do not coincide, reflecting the non-congruent nature of a first-order phase transition with multiple globally conserved charges~\cite{Greiner:1987tg,Glendenning:1992vb,Hempel:2013tfa,Poberezhnyuk:2018mwt}.

Qualitatively, susceptibilities show the expected universal features in the vicinity of the CP~\cite{Stephanov:2008qz,Stephanov:2011pb,Vovchenko:2015pya,Chen:2015dra,Karthein:2025hvl}, including the divergence of $\chi_2^B$, the sign change of $\chi_3^B/\chi_1^B$,
and the peak-and-dip structure of $\chi_4^B/\chi_2^B$.
Negative values of $\chi_3^B/\chi_1^B$ and $\chi_4^B/\chi_2^B$ are also visible below the first-order region at $\mu_B \gtrsim 700$~MeV.
This behavior is associated with the remnants of the nuclear liquid-gas transition contained in the QvdW-HRG model~\cite{Vovchenko:2016rkn}.
Our construction implicitly includes traces of the liquid-gas transition through the QvdW-HRG integration constant for the pressure at $T = 80$~MeV.

The results in Fig.~\ref{fig:hic_contours} indicate that strong remnants of the CP signal should be visible in the equilibrium susceptibilities along the freeze-out line.
This sensitivity is explicitly shown in Fig.~\ref{fig:freezeout_ratios}, where $\chi_2^B$, $\chi_3^B/\chi_1^B$, and $\chi_4^B/\chi_2^B$ are plotted as functions of $\sNN$ along the two freeze-out lines considered.
However, the experiments do not measure equilibrium baryon number fluctuations directly.
Instead, fluctuations of (net-)proton number are measured as a proxy for fluctuations of baryon number, and the results are additionally distorted by non-equilibrium effects, as well as kinematic acceptance, baryon number conservation, centrality selection, and finite-size effects~(see e.g.~\cite{Koch:2025cog} for a review).
Recently, a new method was developed to extract baryon number susceptibilities from net-proton cumulants accounting for most of these effects~\cite{Pihan:2026xng}.
We display the susceptibilities $\chi_2^B$ and $\chi_3^B/\chi_1^B$ extracted in Ref.~\cite{Pihan:2026xng} by symbols in Fig.~\ref{fig:freezeout_ratios} and compare them with the predictions of our EoS.

The left panel of Fig.~\ref{fig:freezeout_ratios} shows $\chi_2^B$ along the freeze-out line.
As the collision energy decreases and the freeze-out line bends toward larger $\mu_B$, approaching the CP, $\chi_2^B$ develops a slight enhancement on both freezeout-lines. 
Our results are in excellent quantitative agreement with the hydrodynamics-based extraction of Ref.~\cite{Pihan:2026xng} at $\sNN \gtrsim 11$~GeV.
At lower energies~($\sNN = 7.7$, $9.2$~GeV), we find a clear deviation where the extracted susceptibilities are enhanced relative to our prediction.
This tension might signal the deterioration of our expansion at larger $\mu_B$, at least on the quantitative level. 
However, dynamical effects important at low energies, but not incorporated in the hydrodynamics-based analysis of~\cite{Pihan:2026xng}, such as memory effects or baryon stopping fluctuations, may also be at play.

The middle panel shows the ratio $\chi_3^B/\chi_1^B$, which is in agreement with the results of Ref.~\cite{Pihan:2026xng} over the range of collision energies considered, albeit with large error bars in the latter.
Toward lower $\sqrt{s_{NN}}$ we observe a clear rise in this ratio in our calculations, driven by the proximity of the freeze-out line to the CP. 
Similarly to the behavior of $\chi_2^B$, the freeze-out line of Ref.~\cite{Vovchenko:2015idt} yields a higher peak than the one from~\cite{Andronic:2017pug}.
This is reflecting the fact that the former freeze-out line passes closer to the CP, as can be seen in Fig.~\ref{fig:HIC_FOPT}. 

Finally, the right panel presents $\chi_4^B/\chi_2^B$, where both freeze-out lines exhibit a similar non-monotonic behavior at low collision energies, including a characteristic change of sign as the freeze-out surface traverses close to the CP. The non-monotonic behavior at low collision energies is qualitatively consistent with predictions from QCD functional renormalization group calculations~\cite{Fu:2023lcm,Zhao:2026mcp} and Dyson--Schwinger studies~\cite{Lu:2025cls}.
We do not overlay the hydrodynamics-based extraction for this ratio, since the corresponding data of Ref.~\cite{Pihan:2026xng} carry uncertainties that are presently too large for a meaningful comparison.

\subsubsection{Speed of sound}

\begin{figure*}
    \centering
    \includegraphics[width=\textwidth]
        {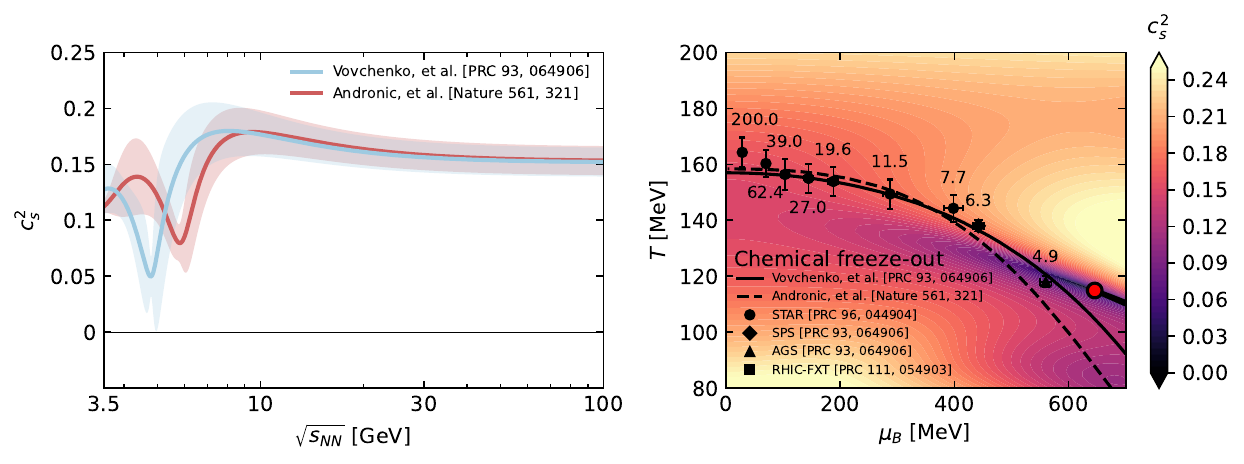}
    \caption{The left panel illustrates the square of the speed of sound along the freeze-out line for heavy-ion collisions. The red band showcases the speed of sound squared on the freeze-out line from Ref.~\cite{Andronic:2017pug} while the blue band corresponds to the freeze-out line from Ref.~\cite{Vovchenko:2015idt}. The softened dip indicates the presence of the CP in the equation of state. The right plot exhibits contours of constant speed of sound squared across the $T-\mu_B$ plane for the heavy-ion collision conditions.  }
    \label{fig:freezeout_cs2}
\end{figure*}
Using the constant entropy-contour expansion, one can calculate the speed of sound by utilizing Eq.~\eqref{eq:cs2_final} in the case of heavy-ion conditions, as shown in Fig.~\ref{fig:freezeout_cs2}. 
The left panel in Fig.~\ref{fig:freezeout_cs2} displays the speed of sound along the freeze-out line from Ref.~\cite{Vovchenko:2015idt} together with the one from Ref.~\cite{Andronic:2017pug}, shown in blue and red, respectively. 
Due to the presence of a CP in the equation of state, we observe a dip as a function of collision energy along both freeze-out lines.
The dip reflects the softening of the EoS near the CP.
Its softness and magnitude depend on the distance between the freeze-out line and the CP. 
This is illustrated in the right panel of Fig.~\ref{fig:freezeout_cs2}, where we show the contour map of $c_s^2$ as a function of $T$ and $\mu_B$ only for the mean value of the parametrization. 
We do note that the exact behavior of $c_s^2$ near the CP depends on the universality class.
For the 3D-Ising class, the speed of sound vanishes at the CP, while in the mean-field universality class, which our expansion belongs to, the value is generally small but non-zero~\cite{folk1999critical,onuki2002phase}.
Given that the freeze-out line never reaches the CP itself, we do not expect the difference between 3D-Ising and mean-field universality classes to have a sizable impact on results in Fig.~\ref{fig:freezeout_cs2}.

\subsubsection{Isentropic trajectories}

\begin{figure}
    \centering
    \includegraphics[width=\columnwidth]
        {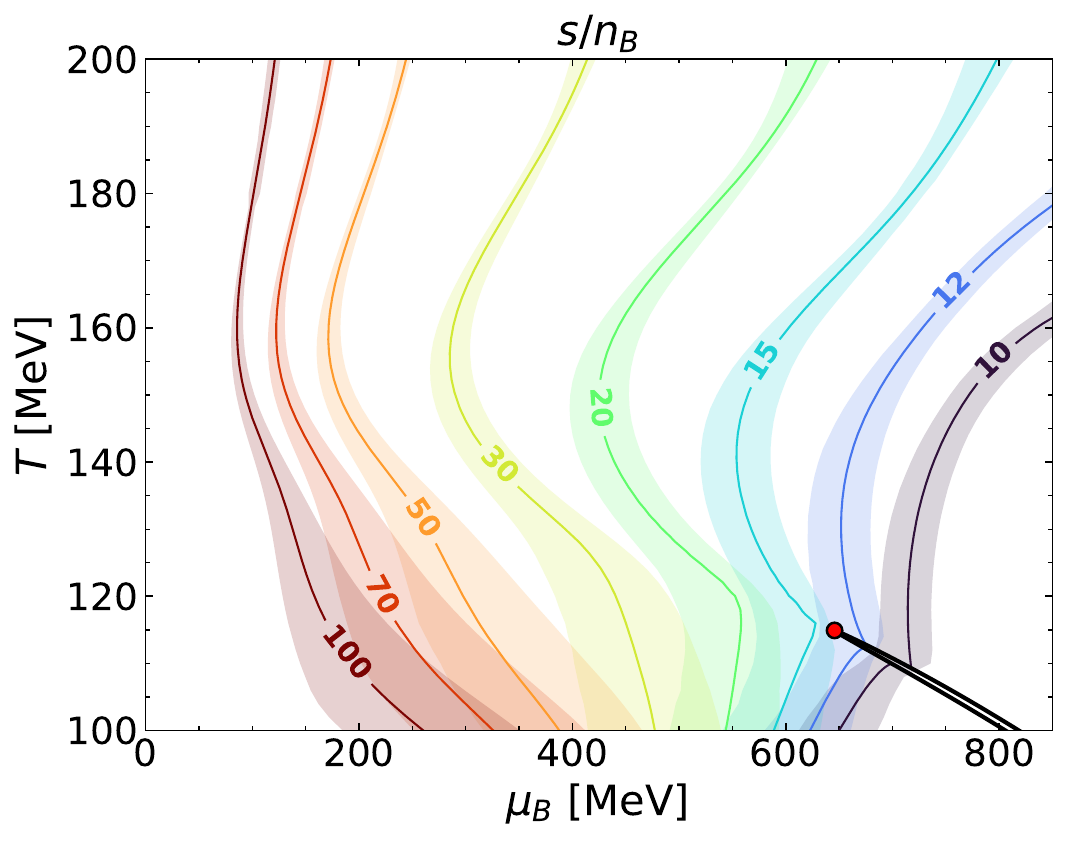}
    \caption{Trajectories of constant entropy per baryon density across the $T-\mu_B$ plane for heavy-ion collisions. The solid curves show selected values of $s/n_B$, while the shaded regions denote the propagated uncertainties.
    }
    \label{fig:hic_isentropes}
\end{figure}

We also calculate isentropic trajectories of constant $s/n_B$ in the heavy-ion $(T,\mu_B)$ plane by calculating the entropy and baryon density with the constraints given in Eq.~\eqref{eq:hic_conditions}. 
Isentropic trajectories are relevant for heavy-ion collisions because, when viscous effects are neglected, the system created in the collision follows constant $s/n_B$ trajectories. Fig.~\ref{fig:hic_isentropes} shows the isentropes in the $(T,\mu_B)$ plane for a number of different values of $s/n_B$. 
At low baryon chemical potential, the trajectories closely resemble those obtained within previous lattice-based equations of state \cite{Gunther:2017sxn} under HIC conditions reflecting the agreement between different expansions at low densities.
As the baryon chemical potential increases, the trajectories bend toward the CP before developing a discontinuity upon crossing the first-order transition line.

Near the phase transition, the entropy and baryon densities become increasingly sensitive to changes in temperature and baryon chemical potential. 
As a result, contours of constant entropy per baryon density tend to focus near the CP, and form a jump across the coexistence region.
These qualitative features are consistent with the two-dimensional equation of state in Ref.~\cite{Shah:2026pue}, which corresponds to the $\mu_Q=\mu_S=0$ plane, while the inclusion of the heavy-ion constraints shifts the lensing of isentropic trajectories toward the CP and first-order transition line determined by the HIC direction. 
We note that we observe both incoming and outgoing isentropes with respect to the first-order region. 
This is qualitatively different from a liquid-gas transition, where only incoming isentropes are present, and consistent with the expectations for the QCD transition~\cite{Wunderlich:2016aed,Pradeep:2024cca}. 

\subsection{Cosmology \& astrophysics}
\label{sec:astro}
\subsubsection{Cosmological QCD phase transition}
In the early Universe, when the temperature dropped to $T\sim 150\, \mathrm{MeV}$, the Quark Gluon Plasma underwent the transition to the confined hadronic phase. While historically it was thought to be a smooth crossover, scenarios with non-vanishing lepton asymmetries can alter this picture, potentially inducing a first-order phase transition~\cite{Schwarz:2009ii}. 
This possibility is particularly compelling in light of the nano-Hertz stochastic gravitational wave background detected by Pulsar Timing Arrays~\cite{Afzal_2023}, which naturally matches the horizon scale at the QCD epoch.
A detailed study of the cosmological QCD phase transition
requires calculating the cosmic trajectory through the three-dimensional QCD phase diagram $(T,\mu_B,\mu_Q)$ and determining whether it intersects the first-order phase transition (FOPT) region. Such calculations have been performed in previous studies using lattice QCD-based EoSs \cite{Oldengott:2017tzj, Wygas:2018otj,Middeldorf-Wygas:2020glx,DiClemente:2025awt,Formaggio:2025nde}. 
However, these studies did not incorporate a CP and a FOPT surface in the three-dimensional phase diagram. The constant-entropy-contour EoS developed in this work allows us, for the first time, to perform a self-consistent analysis of cosmological QCD phase transitions by combining the calculation of cosmic trajectories with a three-dimensional description of the critical structure and the FOPT region.

\begin{figure}[h]
    \centering
    {\includegraphics[width=0.7\linewidth]{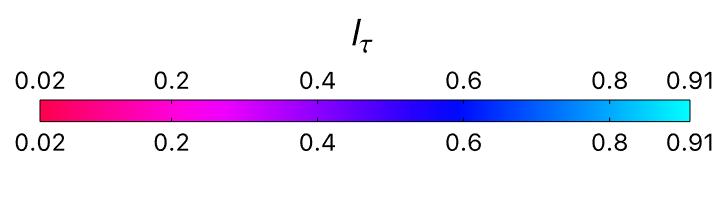}}

    \includegraphics[width=1.\columnwidth,trim={0.5cm 2.8cm 0.5cm 5.1cm},clip]{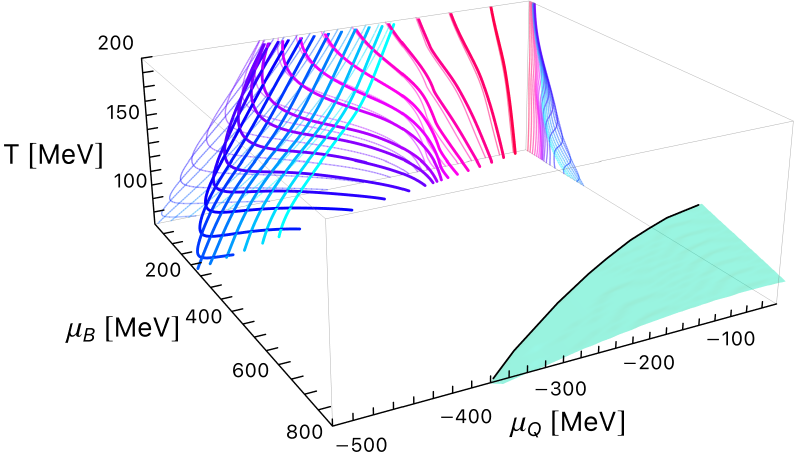}

    \caption{Cosmic trajectories calculated with the constant entropy contour equation of state, for $\ell_e=0$, $\ell_\mu=-\ell_\tau$, and $0.01\leq \ell_\tau\leq 0.91$, compared with the first-order phase transition surface in the $\mu_B-\mu_Q-T$ space. 
    }
    \label{fig:trajectories}
\end{figure}

To calculate the cosmic trajectory we need to solve a system of five conservation equations: baryon asymmetry, electric charge neutrality, and the three individual lepton flavor asymmetries
\begin{align}
    b&=\sum_i B_i \frac{\bar{n}_i }{s}\\
    0&=q =\sum_i Q_i \frac{\bar{n}_i}{s} \\
\ell_j&=\frac{\bar{n}_j+\bar{n}_{\nu_j}}{s}\quad j=\{e,\mu,\tau\}
\end{align}
where $b$ is the baryon asymmetry, $B_i$ and $Q_i$ are the baryon and electric charge fractions of each species, $s$ is the total entropy density and $\ell_j$ are the three lepton asymmetries. The baryon asymmetry is measured to be $b=8.65\times10^{-11}$ \cite{refId0}, whereas only upper bounds exist for the lepton asymmetries: $|\ell_e|\leq10^{-3}$ and $|\ell|\equiv|\ell_e+\ell_\mu+\ell_\tau|\leq10^{-2}$ \cite{Oldengott:2017tzj}. These constraints arise from Big Bang Nucleosynthesis (BBN), which is sensitive to the electron-neutrino asymmetry, and Cosmic Microwave Background (CMB) observations, which constrain the total lepton asymmetry through its contribution to the neutrino energy density and, consequently, the cosmic expansion rate. Within these bounds, the individual lepton flavor asymmetries can be treated as free parameters. In particular, in Fig.~\ref{fig:trajectories}, 
we consider the asymmetric configuration $\ell_e=0$, $\ell_\mu=-\ell_\tau$, which allows us to explore large individual lepton asymmetries while keeping the total lepton asymmetry zero.

Previous works indicated that lepton asymmetries do not push the trajectories to sufficiently high baryon chemical potentials where a FOPT can be expected, while staying within observational constraints~\cite{Oldengott:2017tzj, Wygas:2018otj,Middeldorf-Wygas:2020glx,DiClemente:2025awt,Formaggio:2025nde}.
In Fig.~\ref{fig:trajectories} we show that even when the effect of the charge chemical potential on the CP is taken into account, the transition stays in the crossover regime, with the trajectories remaining below $\mu_B\lesssim 200\,\mathrm{MeV}$ and $|\mu_Q|\lesssim500\,\mathrm{MeV}$ for $T\geq  T_{\mathrm{low}}=80\,\mathrm{MeV}$, confirming that even when very large individual lepton asymmetries are taken into account, the transition is a crossover.
 
It is worth noting that the large values of the electric charge chemical potential reached by our trajectories may make pion condensation relevant, since $|\mu_Q|$ can easily reach the pion mass scale \cite{Vovchenko:2020crk,Floerchinger:2015efa, Ferreira:2025zeu,Ferreira:2026uzi}.
The QvdW-HRG model used as the integration constant includes an effective-mass model for pion condensation, whose contribution is propagated in the construction of the equation of state.
However, as discussed in the beginning of Sec.~\ref{sec:results}, the constant entropy density contour expansion does not explicitly describe the condensation transition.

\subsubsection{Compact stars}
Compact stars and their mergers probe the high-density, low-temperature part of the phase diagram. Strangeness is not conserved in weak interactions, therefore weak equilibrium implies $\mu_S=0$. The charge fraction $Y_Q\equiv n_Q/n_B$, which is equal to the net charged-lepton fraction in neutral matter, typically ranges from 0.05 to 0.1 in cold, $\beta$-equilibrated cores and can reach about 0.3 in proto-neutron stars and in the shocked regions of the merger remnants \cite{Oertel:2016bki,Perego:2019adq}. These conditions select a different section of the four-dimensional equation of state from the heavy-ion surface discussed in Sec.~\ref{sec:hic_results} \cite{Dexheimer:2020zzs,MUSES:2023hyz}. Cold stars, proto-neutron stars and most of the merger remnant remain below the lower limit of the present construction $T_\text{low}=80$ MeV \cite{Perego:2019adq,Radice:2023zlw,Most:2021ktk}. Its direct application is therefore limited to hotter merger regions. These include small hot spots that form at the point of contact when the two cores collide, at densities of about $1$ to $2\,n_{\mathrm{sat}}$ during the first milliseconds \cite{Bauswein:2010dn,Perego:2019adq,Vijayan:2023qrt}, where the maximum temperature can be higher for more compact stars \cite{Most:2021ktk}.

\begin{figure}
    \centering
    \includegraphics[width=\linewidth]
        {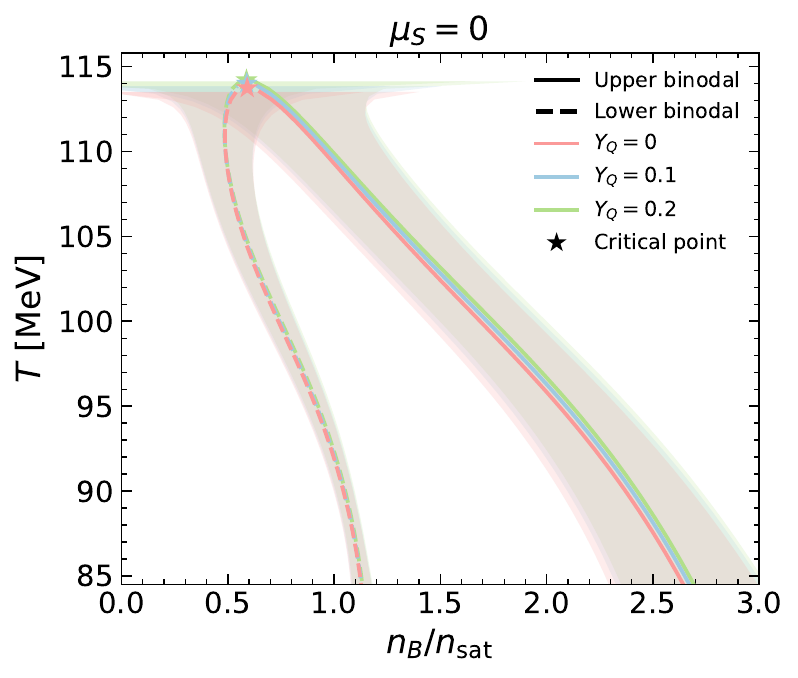}
    \caption{Phase coexistence in the $(T,n_B/n_\mathrm{sat})$ plane at $\mu_S=0$ for fixed total charge fraction $Y_Q=n_Q/nB=0$, $0.1$ and $0.2$. The solid and dashed curves are respectively the higher-density and lower-density binodals. For each $Y_Q$, the mixed phase lies between these boundaries, which meet the CP marked by a star.
    }
    \label{fig:nB_coexistience}
\end{figure}

To examine the phase structure under these conditions, we use the Gibbs construction with conserved baryon number and electric charge. Fig.~\ref{fig:nB_coexistience} shows the two binodals that bound the mixed phase in the ($T,\mu_B$) plane, at $\mu_S=0$, for fixed charge fraction $Y_Q=0$, $0.1$ and $0.2$. These boundaries depend only weakly on $Y_Q$ over this range. They meet near $T_c\simeq114$ MeV and $n_B\simeq0.6\,n_{\mathrm{sat}}$, and separate as the temperature decreases.  At $T=85$ MeV the mixed phase extends from approximately $1.1$ to $2.7\,n_{\mathrm{sat}}$. Within the present extrapolation, this gives a benchmark for comparison with equations of state of models that include a first-order phase transition and are used in merger and supernova simulations \cite{Most:2018eaw,Bauswein:2018bma,Weih:2019xvw,Fischer:2017lag}.

The hot contact regions may reach temperatures and densities within this coexistence region. In less compact binaries, their temperature probably remains below $100$ MeV while their densities can overlap with the predicted mixed phase. In more compact systems, temperatures may locally exceed $T_c$ and matter may encounter the coexistence region as it subsequently cools and decompresses~\cite{Bauswein:2010dn, Most:2021ktk}. 
Whether it crosses a phase boundary or not depends on the trajectory through the phase diagram. This possibility motivates simulations with this EoS to investigate how the transition affects the pressure of the material at the contact interface. Such simulations may also test how phase conversion changes the thermal evolution of the hot streams that feed the disk and the early outflows~\cite{Perego:2019adq}.

The peak temperatures remain sensitive to numerical resolution \cite{Vijayan:2023qrt} and the truncation error up to $\mu_B/T\sim8$ has not been quantified. Although the simulation evolution is slow compared to microscopic strong interactions, the time needed for phase conversion needs to be assessed. The present EoS could supply the high-temperature sector of such simulations, provided that it is thermodynamically consistently matched to a suitable low-temperature EoS, allowing the effects of the predicted phase structure to be investigated while retaining $T_{\mathrm{low}}=80$ MeV as the lower limit of the current construction, similarly to what was done in the $(T,\mu_B)$ plane in Ref. \cite{Yang:2026brr}.

\section{Conclusions}
\label{sec:conclusions}

In this work, we construct a four-dimensional lattice-QCD based equation of state in the $(T,\mus)$ plane using the expansion of constant entropy density contours. By switching the chemical-potential space from Cartesian to spherical coordinates $(\mu,\theta,\varphi)$, each direction reduces to a 1D extrapolation in $\mu$ at fixed $(\theta,\varphi)$, and varying the angles
spans the full 4D space. Since the constant entropy density contour method can incorporate a CP in the expansion, our EoS thus features a critical surface and first-order phase structure. At low chemical potentials $(\mu_B/T\leq3.5)$ we subsequently found agreement between our equation of state and that of recent lattice-based $T^\prime$-expansion scheme results under SN conditions. 

Using our four-dimensional EoS, we can fix the densities $n_s=0,~n_Q/n_B=0.4$ to obtain a two-dimensional EoS for the conditions found in heavy-ion collisions. 
In this particular setup, the CP shifts to $(T,\mu_B)=(115\pm7,651\pm74)$ MeV at expansion $\mathcal{O}(\mu^2)$. Hence we find the collision energy $\sqrt{s_{NN}}$ between $3-5.2$ GeV to be in the vicinity of CP in the heavy-ion plane. Taking the $\mathcal{O}(\mu^2)$ expansion as exact, we reconstructed the full pressure $P(T,\boldsymbol{\mu})$ under these conditions, from which all thermodynamic quantities and their higher-order derivatives follow. Using the entropy density and baryon density, we studied the behavior
of isentropic trajectories and their focusing in the vicinity of the CP. Additionally, the speed of sound squared has been calculated across the $T-\mu_B$ plane and on the freeze-out lines. A soft dip was found on the freeze-out lines, due to the presence of the CP in the equation of state.

Ratios of baryon cumulants were calculated on the chemical freeze-out from the equation of state, and we found that both the ratios  $\chi_4^B/\chi_2^B$ and $\chi_3^B/\chi_1^B$ diverge at the CP. The strength and location of the resulting non-monotonic signatures depend on the distance of the freeze-out curve from to the CP, which confirms the findings of Refs. \cite{Mroczek:2020rpm,Karthein:2025hvl}. Subsequently, we compared the baryon cumulants from our EoS with the baryon cumulants extracted from net proton measurements using the hydrodynamics-based approach \cite{Pihan:2026xng} and found agreement for $\chi_3^B/\chi_1^B$ on the freeze-out line. For $\chi_2^B$ on the freeze-out line we found a discrepancy, which might be due to the systematic error from truncating our expansion to second order, or maybe to retardation effects related to finite relaxation time for critical fluctuations in the hydrodynamics-based approach.

We also showed results for the four-dimensional equation of state in relation to the early Universe and to the conditions relevant to neutron-star mergers. The cosmological trajectories considered here do not intersect the first-order surface, even when large lepton flavor asymmetries are considered. Under compact-star conditions, $\mu_S=0$, the coexistence region varies only weakly with $Y_Q$ in the range $0\leq Y_Q\leq0.2$, the CPs lying near $T_c\simeq114$ MeV and $n_B\simeq0.6\,n_{\mathrm{sat}}$. Within the scope of the present extrapolation, this region may overlap the conditions reached in merger hot spots, providing a lattice-informed benchmark for studies of the effects of the phase transition. The EoS presented here could be used to supply the high-temperature part of merger simulations by means of a thermodynamically consistent matching to an appropriate low-temperature EoS, enabling the astrophysical implications to be examined.

In the future, this 4D equation of state with a critical surface can be adapted in relativistic hydrodynamic simulations of heavy-ion collisions. Coupled to a dynamical description of critical fluctuations, this would make it possible to compute critical signatures beyond the equilibrium approximation used here, accounting for the finite relaxation time of critical modes, the evolution along realistic trajectories, and particlization at freeze-out. Furthermore, all results presented here were obtained by truncating the constant-entropy-contour expansion at $\mathcal{O}(\mu^2)$. Including the next order in the expansion, once the corresponding lattice coefficients become available with sufficient precision, will allow us to systematically quantify the truncation uncertainty and assess the robustness of the critical point location and of the associated fluctuation observables. 

\section*{Acknowledgements}
This material is based upon work supported by the National Science Foundation under grants No. PHY-2208724, PHY-2116686, PHY-2514763, PHY-2621752 and PHY-2623480, and within the framework of the MUSES collaboration, under Grant No. OAC-2103680. This material is also based upon work supported by the U.S. Department of Energy, Office of Science, Office of Nuclear Physics, under Award Number DE-SC0022023, as well as by the National Aeronautics and Space Administration (NASA) under Award Number 80NSSC24K0767. 
VV was supported by the U.S. Department of Energy, Office of Science, Office of Nuclear Physics, Early Career Research Program under Award No.~DE-SC0026065.
MH is supported by the Brazilian National Council for Scientific and Technological Development (CNPq) under process No. 313638/2025-0 and acknowledges the
support of the INCT-FNA grant 408419/2024-5. HS
is also supported by the Deutsche Forschungsgemeinschaft
(DFG, German Research Foundation) through the CRC-
TR 211 “Strong-interaction matter under extreme conditions” Project No. 315477589–TRR 211.
% \clearpage
\bibliography{main}

\clearpage

\widetext
\begin{appendix}
\widetext 
\setcounter{equation}{0}
\setcounter{figure}{0}
\renewcommand{\theequation}{A.\arabic{equation}}
\renewcommand{\thefigure}{A.\arabic{figure}}
\makeatletter
\section*{SUPPLEMENTAL MATERIAL}
\appendix

\section{Additional cumulant ratios across the HIC phase diagram}

\subsection{ Electric and strangeness charge cumulant ratios on the freeze-out line}
In this section, we present additional conserved-charge observables evaluated along the chemical freeze-out trajectories discussed in Sec.~\ref{sec:hic_results}.

\begin{figure*}[b]
    \centering
    \includegraphics[width=\textwidth]
        {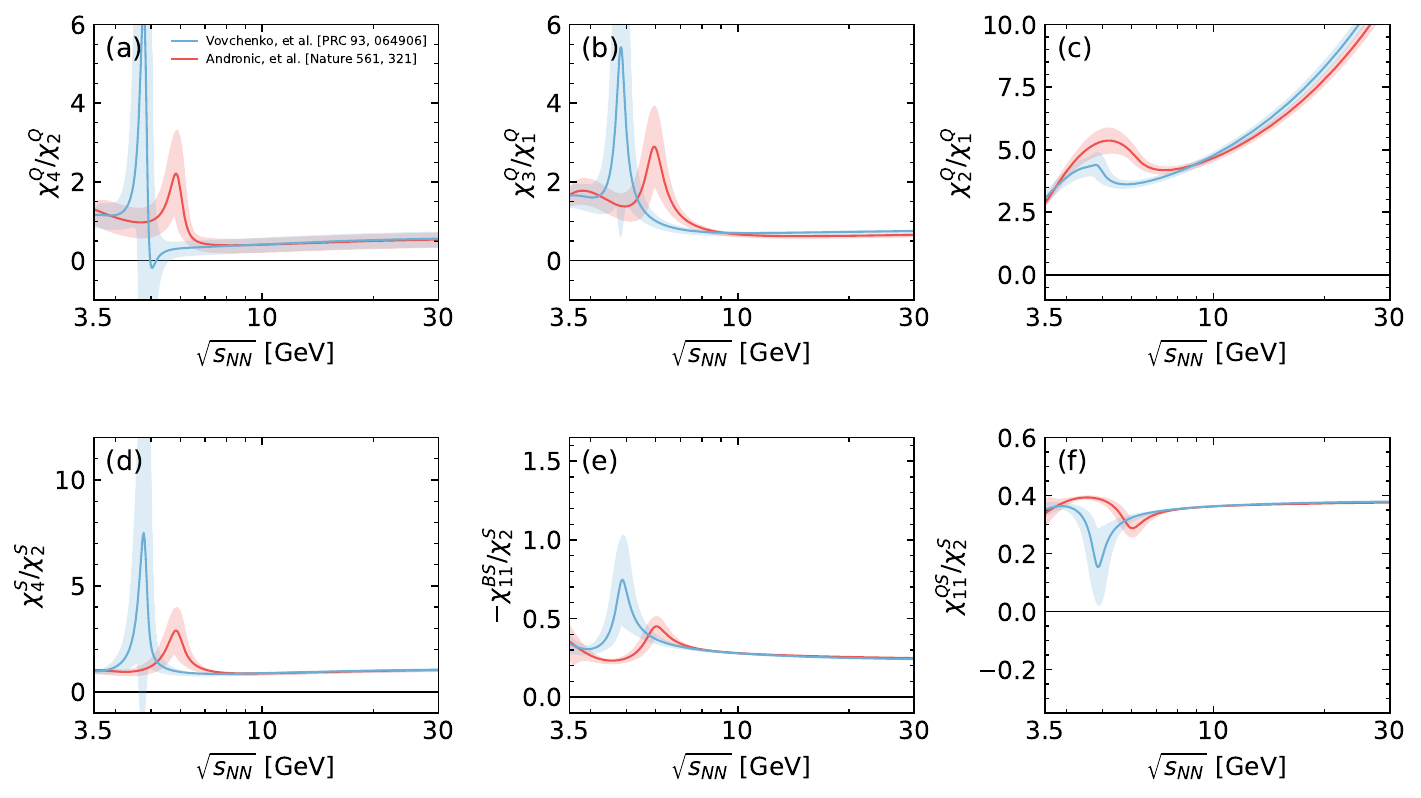}
    \caption{The temperature dependence of additional susceptibility and chemical-potential ratios on the chemical freeze-out lines in the heavy-ion condition plane. These include (a), $\chi_4^Q/\chi_2^Q$, (b) $\chi_3^Q/\chi_1^Q$, (c) $\chi_2^Q/\chi_1^Q$, (d) $\chi_4^S/\chi_2^S$, (e) $-\chi_{11}^{BS}/\chi_2^S$, and (f) $\chi_{11}^{QS}/\chi_2^S$.
    }
    \label{fig:appendix_composite}
\end{figure*}

The top panel of Fig.~\ref{fig:appendix_composite} displays the electric charge cumulant ratios evaluated along the same two freeze-out curves. Similar to its baryon number counterpart, at high $\sqrt{s_{NN}}$ the electric charge cumulants show a monotonic behavior. As the collision energy decreases, we again observe a non-monotonic diverging behavior for $\chi_4^Q/\chi_2^Q$ and $\chi_3^Q/\chi_1^Q$.
At low collision energies, these ratios peak, albeit at a lower magnitude than the corresponding baryon-number ratios.
The suppressed peaks are a direct consequence
of the low values of $\mu_Q$ under the heavy-ion conditions $n_S=0,n_Q/n_B=0.4$. 

The bottom left plot of Fig.~\ref{fig:appendix_composite} shows the strangeness susceptibility ratio $\chi_4^S/\chi_2^S$ along the two chemical freeze-out parametrizations. This ratio's behavior is similar to that of the electric charge ratios, with sharper peaks indicative of the larger value of $\mu_S$ under these conditions. Heavy-ion conditions $\chi_1^S=0$ require that we omit the third and second strangeness cumulant ratios.

We present the cross-covariances between fluctuations of more than one conserved charge by calculating $-\chi_{11}^{BS}/\chi_2^S$ and $\chi_{11}^{QS}/\chi_2^S$ along the freeze-out lines in the bottom-mid and bottom-right panels Fig~\ref{fig:appendix_composite} respectively. We observe a soft signal of criticality from the equation of state by the soft dips in the mixed conserved-charge correlations at low collision energy.

\begin{figure}[t]
    \centering
    \includegraphics[width=0.5\textwidth]
        {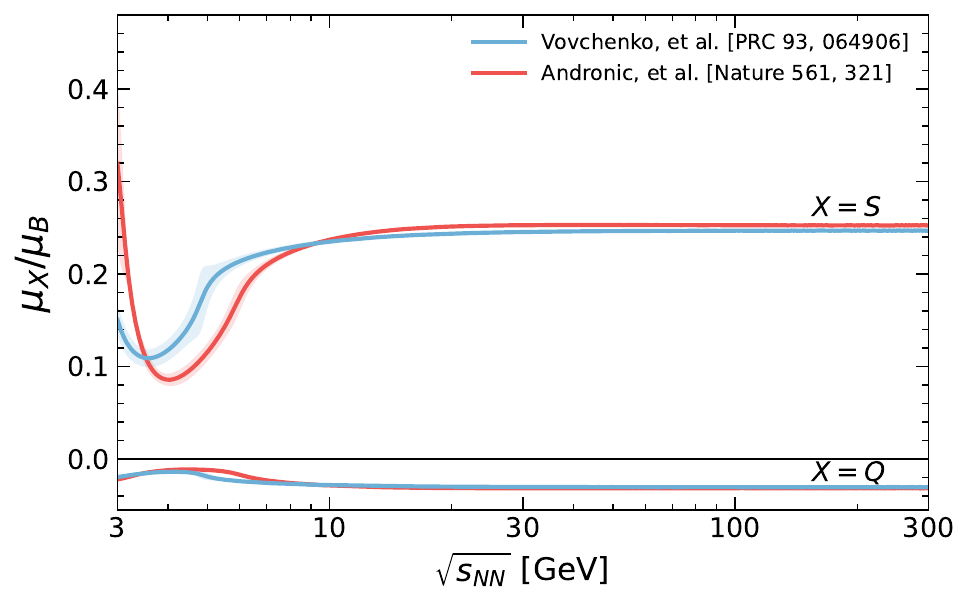}
    \caption{Chemical-potential ratios $(\mu_S/\mu_B,\mu_Q/\mu_B)$ along both freeze-out lines under heavy-ion conditions ($n_s=0,n_Q/n_B=0.4)$.%\MH{Add et al. to Andronic and replace VBG with Vovchenko et al. to make the legend consistent with the rest of the paper.}
    }
    \label{fig:mu_ratios_freezeout}
\end{figure}

Figure~\ref{fig:mu_ratios_freezeout} shows the ratios $\mu_S/\mu_B$ and $\mu_Q/\mu_B$ along the two chemical freeze-out parametrizations. Both freeze-out curves show a similar qualitative dependence of $\mu_Q$ on the collision energy, when the heavy-ion conditions are implemented. A similar behavior was observed in Figure \ref{fig:muSovermuB} of the main text. We note that the absolute values of both $\mu_S/\mu_B$ and $\mu_Q/\mu_B$ have minima at $\sqrt{s_{NN}}\approx 4$ GeV. Both ratios approach a nearly constant value of $\mu_S/\mu_B\approx0.25$, consistent with the behavior expected from strangeness neutrality at small baryon chemical potential \cite{Borsanyi:2020fev}.

\subsection{ Temperature derivatives across the phase diagram}
In Refs.~\cite{Chen:2025vwl,Mallick:2026dsc}, the following  temperature derivatives were suggested to be comparable to temperature cumulants. These susceptibility ratios are defined below:
\begin{align}
\label{eq:T_sus}
\tau_2&=\frac{1}{\chi_2^{T}},\qquad
\tau_3=-\frac{\chi_3^{T}}{(\chi_2^{T})^3},\qquad
\tau_4=\frac{3(\chi_3^{T})^2}{(\chi_2^{T})^5}-\frac{\chi_4^{T}}{(\chi_2^{T})^4},
\end{align}
where the $\chi^T_n$ are defined as:
\begin{equation}
\chi_n^{T}(T,\mu_B,\mu_Q,\mu_S)=T^{\,n-4}\left(\frac{\partial^n p}{\partial T^n}\right)_{\mu_B,\mu_Q,\mu_S},
\end{equation}
In this work, all temperature susceptibilities are evaluated at fixed $(\mu_B,\mu_Q,\mu_S)$.
 
\begin{figure*}
    \centering
    \includegraphics[width=\textwidth]
        {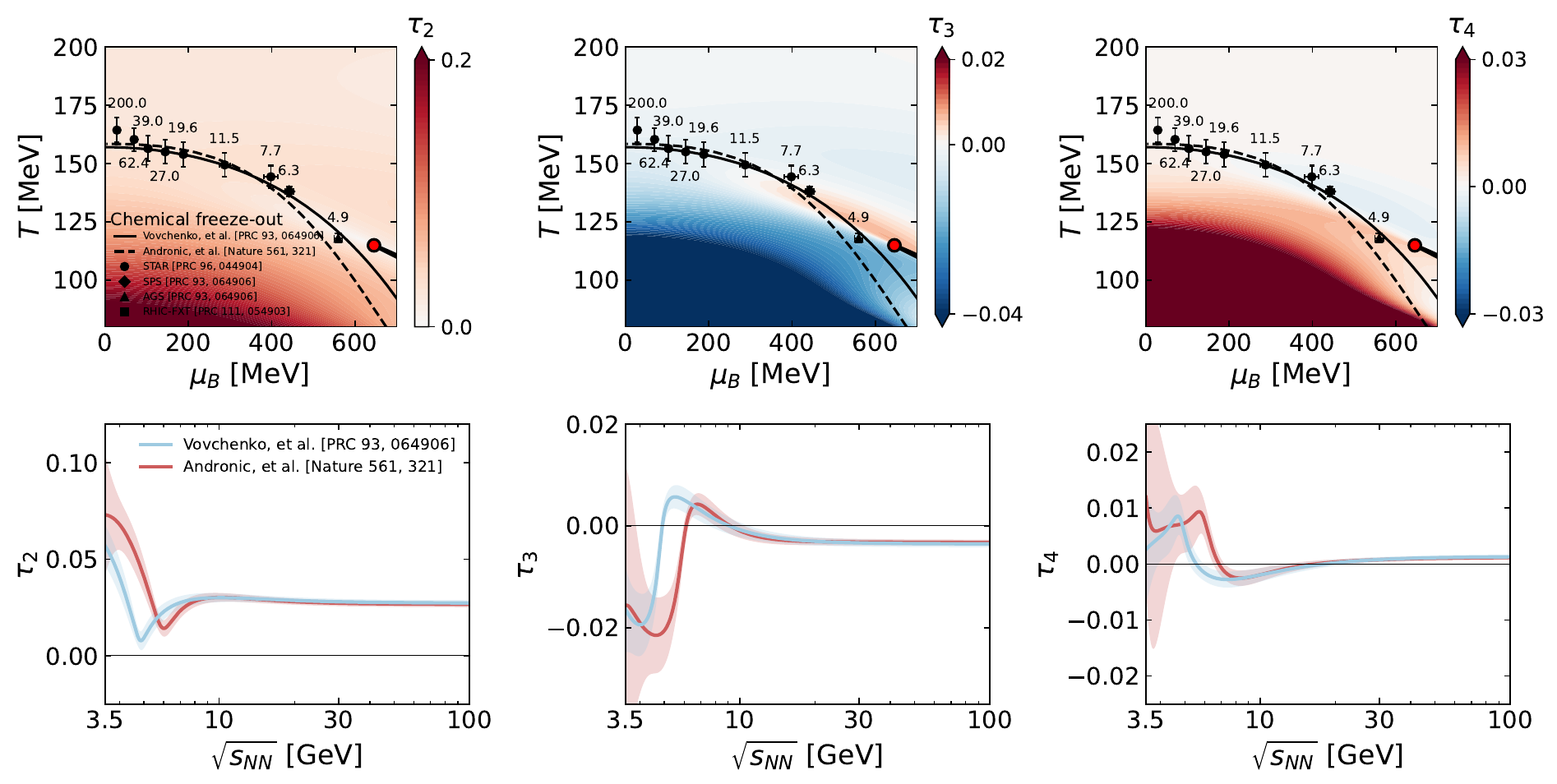}
    \caption{The top panel in the figure shows contours of ratios of temperature derivatives $\tau_2$, $\tau_3$, and $\tau_4$ from the left to right respectively, across the $T$--$\mu_B$ plane, under heavy ion conditions. The bottom panel displays the same ratios of temperature derivatives as a function of collision energy calculated on two different freeze-out lines from Refs.~\cite{Andronic:2017pug} and~\cite{Vovchenko:2015idt} shown in red and blue respectively. }
    \label{fig:t_cumulants}
\end{figure*}

In Fig.~\ref{fig:t_cumulants}, we calculate the temperature susceptibilities across the $T$--$\mu_B$ plane under HIC conditions. The top panel of Fig.~\ref{fig:t_cumulants} shows contour plots of $\tau_2$, $\tau_3$, and $\tau_4$, from left to right.
As the critical point is approached, $\tau_2$ tends to zero. This follows directly from $\tau_2 = 1/\chi_2^{T}$: the heat capacity at fixed chemical potentials grows near the CP, so the variance of the temperature is suppressed.
The higher-order temperature derivatives $\tau_3$ and $\tau_4$ do not show the critical scaling structure seen in the higher-order baryon-number cumulants (Fig.~\ref{fig:hic_contours} in the main text). 

The bottom panel of Fig.~\ref{fig:t_cumulants} shows $\tau_2$, $\tau_3$, and $\tau_4$ along the two chemical freeze-out lines. Along the freeze-out lines of Refs.~\cite{Andronic:2017pug} and~\cite{Vovchenko:2015idt}, shown in red and blue respectively, $\tau_2$ develops a dip as a function of collision energy, consistent with the enhanced heat capacity in the vicinity of the CP of our equation of state. A dip alone, however, is not a unique signature of criticality. Ref.~\cite{Mallick:2026dsc} finds a shallow minimum of $\tau_2$ along the freeze-out curve near $\sqrt{s_{NN}}\simeq 11$~GeV even in the ideal HRG, without any critical point. That minimum arises from the shape of the freeze-out trajectory itself: the freeze-out temperature saturates at high energies while $\mu_B$ keeps growing toward lower energies. Interestingly, the STAR collaboration recently reported a non-monotonic collision-energy dependence of two-particle transverse-momentum correlations in central Au+Au collisions at $\sqrt{s_{NN}}=3.0$--$7.7$~GeV~\cite{STAR:2026vjv} which might be influenced by the critical point.
Although $\langle p_T\rangle$ fluctuations are related to temperature fluctuations of the fireball~\cite{Gavin:2003cb}, the measured signal also contains contributions from initial-state geometry, radial flow, and acceptance effects. Since the $\tau_n$ computed here are equilibrium quantities and also not necessarily related to temperature fluctuations, a direct comparison with the STAR data would require modeling the full collision dynamics, which is beyond the scope of this work.

The middle and right panels of the bottom row of Fig.~\ref{fig:t_cumulants} show $\tau_3$ and $\tau_4$ as functions of $\sqrt{s_{NN}}$. Their behavior differs markedly from the HRG baseline of Ref.~\cite{Mallick:2026dsc}, where $\tau_3$ remains negative and $\tau_4$ positive along the freeze-out curve, with only smooth extrema near $\sqrt{s_{NN}}\simeq 3$~GeV. This suggests that higher-order $\langle p_T\rangle$ fluctuations, such as the skewness and kurtosis already measured at the LHC~\cite{ALICE:2023tej}, could help distinguish critical behavior from a non-critical hadronic baseline. Such measurements would be particularly interesting at low collision energies, e.g., in the STAR fixed-target program, HADES, and the upcoming CBM experiment, where the freeze-out lines approach the critical region of our equation of state.

\section{Derivation of the Equation of State}
\label{app:derivations}
This appendix outlines the derivation of the thermodynamic observables used throughout this work. We begin with the constant-entropy contour expansion and derive the conserved charge densities and susceptibilities through repeated differentiation with respect to the chemical potentials.

\subsection{Constant-Entropy Expansion}

The equation of state is constructed by expanding the constant-entropy contours in the magnitude of the chemical potential,
\begin{equation}
\label{eq:Tso2}
T (T_0,\theta,\varphi,\mu)= T_0 + \frac{1}{2}\alpha_2(T_0,\theta,\varphi)\mu^2,
\end{equation}
where
\begin{equation}
\mu = \sqrt{\mu_B^2+\mu_Q^2+\mu_S^2}.
\end{equation}

Throughout this appendix, $'$ denotes differentiation with respect to $T_0$ at fixed $(\theta,\varphi)$
\begin{equation}
f'(T_0,\theta,\varphi)
\equiv
\left(
\frac{\partial f}{\partial T_0}
\right)_{\theta,\varphi}.
\end{equation}

Partial derivatives with respect to the conserved chemical potentials are evaluated through the chain rule,
\begin{equation}
\frac{\partial}{\partial\mu_i}
=
\frac{\partial\mu}{\partial\mu_i}
\frac{\partial}{\partial\mu}
+
\frac{\partial\theta}{\partial\mu_i}
\frac{\partial}{\partial\theta}
+
\frac{\partial\varphi}{\partial\mu_i}
\frac{\partial}{\partial\varphi},
\label{eq:A_chain}
\end{equation}
where $i=B,Q,S$. The required coordinate derivatives are obtained directly from the definitions of
$\mu$, $\theta$, and $\varphi$ given in Sec.~II and are used throughout the following derivations.

\subsection{The Pressure from Entropy Density}
We calculate the pressure by integrating the entropy density with respect to temperature while holding chemical potentials fixed:
\begin{align}
    s(T,\mus)=\left(\frac{\partial P(T,\mus)}{\partial T}\right)_{\mus},
\end{align}
yielding
\begin{align}
\label{eq:pressure}
    \int_{T\low}^T d \tilde T s(\tilde T,\mus)=P(T,\mus)-P(T\low,\mus),
\end{align}
where we set $T\low=80$ MeV, and calculate $P(T\low,\boldsymbol{\mu})$ using the Quantum van der Waals Hadron Resonance Gas (QvdW-HRG) model.

\subsection{Conserved Charge Densities}
\label{sec:densitiy_derivs}
The conserved charge densities are obtained by differentiating the entropy integral with respect to the corresponding chemical potential,
\begin{equation}
n_i(T,\mu_B,\mu_Q,\mu_S)
=
n_i^{\mathrm{HRG}}(T_{\mathrm{low}},\mu_B,\mu_Q,\mu_S)
+
\left(
\frac{\partial}{\partial\mu_i}
\right)_{T,\mu_{j\neq i}}
\int_{T_{\mathrm{low}}}^{T}
d\tilde T\,
s\!\left[T_0(\tilde T,\mu_B,\mu_Q,\mu_S)\right],
\label{eq:A_density_start}
\end{equation}
where $i\neq j\neq k$, where $\tilde T_0$ is the integration variable, and $s$ has dimensions [MeV]$^3$. 

Applying the chain rule gives
\begin{equation}
\left(
\frac{\partial s}{\partial\mu_i}
\right)_{T,\mu_{j\neq i}}
=
s^\prime
\left(
\frac{\partial T_0}{\partial\mu_i}
\right)_{T,\mu_{j\neq i}},
\label{eq:A_ds}
\end{equation}

where the derivatives of $T_0$ are found by differentiating Eq.\eqref{eq:Tso2},
 with respect to the corresponding chemical potential, while holding the temperature fixed. This yields
\begin{align}
\left(
\frac{\partial T_0}{\partial\mu_B}
\right)_{T,\mu_Q,\mu_S}
&=
\frac{
-\mu_B\alpha_2
+
\frac{\sqrt{\mu_Q^2+\mu_S^2}}{2}
\left(
\frac{\partial\alpha_2}{\partial\theta}
\right)_{T_0,\varphi}
}
{
1+\frac{\mu^2}{2}\alpha_2'
},
\\[1ex]
\left(
\frac{\partial T_0}{\partial\mu_Q}
\right)_{T,\mu_B,\mu_S}
&=
\frac{
-\mu_Q\alpha_2
-
\frac{\mu_B\mu_Q}{2\sqrt{\mu_Q^2+\mu_S^2}}
\left(
\frac{\partial\alpha_2}{\partial\theta}
\right)_{T_0,\varphi}
+
\frac{\mu_S\mu^2}{2\sqrt{\mu_Q^2+\mu_S^2}^2}
\left(
\frac{\partial\alpha_2}{\partial\varphi}
\right)_{T_0,\theta}
}
{
1+\frac{\mu^2}{2}\alpha_2'
},
\\[1ex]
\left(
\frac{\partial T_0}{\partial\mu_S}
\right)_{T,\mu_B,\mu_Q}
&=
\frac{
-\mu_S\alpha_2
-
\frac{\mu_B\mu_S}{2\sqrt{\mu_Q^2+\mu_S^2}}
\left(
\frac{\partial\alpha_2}{\partial\theta}
\right)_{T_0,\varphi}
-
\frac{\mu_Q\mu^2}{2\sqrt{\mu_Q^2+\mu_S^2}^2}
\left(
\frac{\partial\alpha_2}{\partial\varphi}
\right)_{T_0,\theta}
}
{
1+\frac{\mu^2}{2}\alpha_2'
}.
\label{eq:A_dT0}
\end{align}

Substituting Eq.~(\ref{eq:A_dT0}) into Eq.~(\ref{eq:A_ds}), and then into Eq.~(\ref{eq:A_density_start}), we change the integration variable from $T'$ to $T_0'$ using
\begin{equation}
\left(
\frac{dT}{dT_0}
\right)_{\mu,\theta,\varphi}
=
1+\frac{\mu^2}{2}\alpha_2',
\end{equation}
which cancels the denominator in Eq.~(\ref{eq:A_dT0}). The conserved charge densities therefore become
\begin{align}
n_B(T,\mu_B,\mu_Q,\mu_S)
&=
n_B^{\mathrm{HRG}}
+
\int_{T_{0,\mathrm{low}}}^{T_0}
d\tilde T_0\,
s'
\left[
-\mu_B\alpha_2
+
\frac{\sqrt{\mu_Q^2+\mu_S^2}}{2}
\left(
\frac{\partial\alpha_2}{\partial\theta}
\right)_{\tilde T_0,\varphi}
\right],
\\
n_Q(T,\mu_B,\mu_Q,\mu_S)
&=
n_Q^{\mathrm{HRG}}
+
\int_{T_{0,\mathrm{low}}}^{T_0}
d\tilde T_0\,
s'
\left[
-\mu_Q\alpha_2
-
\frac{\mu_B\mu_Q}{2\sqrt{\mu_Q^2+\mu_S^2}}
\left(
\frac{\partial\alpha_2}{\partial\theta}
\right)_{\tilde T_0,\varphi}
+
\frac{\mu_S\mu^2}{2\sqrt{\mu_Q^2+\mu_S^2}^2}
\left(
\frac{\partial\alpha_2}{\partial\varphi}
\right)_{\tilde T_0,\theta}
\right],
\\
n_S(T,\mu_B,\mu_Q,\mu_S)
&=
n_S^{\mathrm{HRG}}
+
\int_{T_{0,\mathrm{low}}}^{T_0}
d\tilde T_0\,
s'
\left[
-\mu_S\alpha_2
-
\frac{\mu_B\mu_S}{2\sqrt{\mu_Q^2+\mu_S^2}}
\left(
\frac{\partial\alpha_2}{\partial\theta}
\right)_{\tilde T_0,\varphi}
-
\frac{\mu_Q\mu^2}{2\sqrt{\mu_Q^2+\mu_S^2}^2}
\left(
\frac{\partial\alpha_2}{\partial\varphi}
\right)_{\tilde T_0,\theta}
\right].
\end{align}

Finally, using
\begin{equation}
\alpha_2(T_0,\theta,\varphi)
=
-
\frac{
[T_0^2 X_2(T_0,\theta,\varphi)]^\prime
}{
s^\prime(T_0)
}
=
-
\frac{
2T_0X_2(T_0,\theta,\varphi)+T_0^2X_2^\prime(T_0,\theta,\varphi)
}{
s^\prime(T_0)
},
\label{eq:A_alpha2}
\end{equation}
where $X_2$ is the second-order generalized susceptibility:
\begin{align}
    X_n^{\theta,\varphi}(T)=\frac{\partial^n(P(T,\mu,\theta,\varphi)/T^4)}{\partial(\mu/T)^n}\Bigg\vert_{\theta,\varphi},
\end{align}
the remaining integrals simplify according to
\begin{equation}
\int_{T_{0,\mathrm{low}}}^{T_0}
d\tilde T_0
\,
s'
\alpha_2
=
-\int_{T_{0,\mathrm{low}}}^{T_0}
d\tilde T_0
\frac{\partial(\tilde T_0^2X_2)}{\partial \tilde T_0}
=
-\tilde T_0^2X_2(\tilde T_0)\boundary,
\label{eq:A_DeltaX2}
\end{equation}
yielding
\begin{align}
n_B(T,\mu_B,\mu_Q,\mu_S)
&=
n_B^{\mathrm{HRG}}
+
\bigg[\mu_B \tilde T_0^2 X_2
-
\frac{\sqrt{\mu_Q^2+\mu_S^2}}{2} \tilde T_0^2 
\left(
\frac{\partial X_2}{\partial\theta}
\right)_{\tilde T_0,\varphi}\bigg]\boundary,
\\
n_Q(T,\mu_B,\mu_Q,\mu_S)
&=
n_Q^{\mathrm{HRG}}
+
\bigg[\mu_Q \tilde T_0^2  X_2
+
\frac{\mu_B\mu_Q}{2\sqrt{\mu_Q^2+\mu_S^2}} \tilde T_0^2 
\left(
\frac{\partial X_2}{\partial\theta}
\right)_{\tilde T_0,\varphi}
-
\frac{\mu_S\mu^2}{2\sqrt{\mu_Q^2+\mu_S^2}^2} \tilde T_0^2 
\left(
\frac{\partial X_2}{\partial\varphi}
\right)_{\tilde T_0,\theta}\bigg]\boundary,
\\
n_S(T,\mu_B,\mu_Q,\mu_S)
&=
n_S^{\mathrm{HRG}}
+
\bigg[\mu_S \tilde T_0^2  X_2
+
\frac{\mu_B\mu_S}{2\sqrt{\mu_Q^2+\mu_S^2}} \tilde T_0^2 
\left(
\frac{\partial X_2}{\partial\theta}
\right)_{\tilde T_0,\varphi}
+
\frac{\mu_Q\mu^2}{2\sqrt{\mu_Q^2+\mu_S^2}^2} \tilde T_0^2 
\left(
\frac{\partial X_2}{\partial\varphi}
\right)_{\tilde T_0,\theta}\bigg]\boundary.
\end{align}

\subsection{Conserved Charge Susceptibilities}
\label{sec:sus_derivs}

The diagonal second-order susceptibilities are obtained by
differentiating each conserved charge density with respect to its
corresponding chemical potential,
\begin{equation}
T^2 \chi_2^i(T,\mu_B,\mu_Q,\mu_S)
=
\left(
\frac{\partial n_i}{\partial\mu_i}
\right)_{T,\mu_{j\neq i}},
\label{eq:A_chi2_definition}
\end{equation}

\subsubsection{Second-order diagonal susceptibilities}

Differentiating the baryon density gives
\begin{align}
T^2 \chi_2^B(T,\mu_B,\mu_Q,\mu_S)
&=
T_\text{low}^2 \chi_{2,\mathrm{HRG}}^B
+
\bigg[ \tilde T_0^2  X_2
+
\mu_B  
\left(
\frac{\partial(\tilde T_0^2 X_2)}{\partial\mu_B}
\right)_{T,\mu_Q,\mu_S}
\nonumber\\
&\quad
-
\frac{\sqrt{\mu_Q^2+\mu_S^2}}{2}
\left(
\frac{\partial}{\partial\mu_B}
\left[
\left(
\frac{\partial(\tilde T_0^2 X_2)}{\partial\theta}
\right)_{\tilde T_0,\varphi}
\right]
\right)_{T,\mu_Q,\mu_S}\bigg]\boundary.
\label{eq:A_chi2B}
\end{align}
The first chemical-potential derivative appearing here is
\begin{align}
\left(
\frac{\partial (T_0^2 X_2)}{\partial\mu_B}
\right)_{T,\mu_Q,\mu_S}
&=
\frac{d}{dT_0}\Biggl[T_0^2 X_2\Biggr]
\left(
\frac{\partial T_0}{\partial\mu_B}
\right)_{T,\mu_Q,\mu_S}
-
\frac{\sqrt{\mu_Q^2+\mu_S^2}}{\mu^2}
\left(
\frac{\partial (T_0^2 X_2)}{\partial\theta}
\right)_{T_0,\varphi}.
\label{eq:A_dDeltaX2_dmuB}
\end{align}
The derivative of the angular term follows in the same manner,
\begin{align}
&
\left(
\frac{\partial}{\partial\mu_B}
\left[
\left(
\frac{\partial (T_0^2 X_2)}{\partial\theta}
\right)_{T_0,\varphi}
\right]
\right)_{T,\mu_Q,\mu_S}
=
\left(
\frac{\partial}{\partial T_0}
\left[
\left(
\frac{\partial (T_0^2 X_2)}{\partial\theta}
\right)_{T_0,\varphi}
\right]
\right)_{\theta,\varphi}
\left(
\frac{\partial T_0}{\partial\mu_B}
\right)_{T,\mu_Q,\mu_S}
-
\frac{\sqrt{\mu_Q^2+\mu_S^2}}{\mu^2}
\left(
\frac{\partial^2 (T_0^2 X_2)}{\partial\theta^2}
\right)_{T_0,\varphi}.
\label{eq:A_dDeltaX2theta_dmuB}
\end{align}

Derivatives of $X_2$ with respect to $\mu_Q$ and $\mu_S$ are as follows:
\begin{align}
    \left(
    \frac{\partial (T_0^2 X_2)}{\partial\mu_Q}
    \right)_{T,\mu_B,\mu_S}
    &=
    \frac{d}{dT_0}\Biggl[T_0^2 X_2\Biggr]
    \left(
    \frac{\partial T_0}{\partial\mu_Q}
    \right)_{T,\mu_B,\mu_S}
    +
    \frac{\mu_B\mu_Q}{\mu^2\sqrt{\mu_Q^2+\mu_S^2}}
    \left(
    \frac{\partial (T_0^2 X_2)}{\partial\theta}
    \right)_{T_0,\varphi}
    -
    \frac{\mu_S}{\mu_Q^S+\mu_S^2}
    \left(
    \frac{\partial (T_0^2 X_2)}{\partial\varphi}
    \right)_{T_0,\theta},
    \\
    \left(
    \frac{\partial (T_0^2 X_2)}{\partial\mu_S}
    \right)_{T,\mu_B,\mu_Q}
    &=
    \frac{d}{dT_0}\Biggl[T_0^2 X_2\Biggr]
    \left(
    \frac{\partial T_0}{\partial\mu_S}
    \right)_{T,\mu_B,\mu_Q}
    +
    \frac{\mu_B\mu_S}{\mu^2\sqrt{\mu_Q^2+\mu_S^2}}
    \left(
    \frac{\partial (T_0^2 X_2)}{\partial\theta}
    \right)_{T_0,\varphi}
    +
    \frac{\mu_Q}{\mu_Q^S+\mu_S^2}
    \left(
    \frac{\partial (T_0^2 X_2)}{\partial\varphi}
    \right)_{T_0,\theta},
\label{eq:A_dDeltaX2_dmuS}
\end{align}
where their angular derivatives are treated in the same manner as in Eq.~(\Ref{eq:A_dDeltaX2theta_dmuB}).

Differentiating the electric-charge density gives
\begin{align}
T^2 \chi_2^Q(T,\mu_B,\mu_Q,\mu_S)
&=
T_\text{low}^2 \chi_{2,\mathrm{HRG}}^Q
+
\bigg[ \tilde T_0^2  X_2
+
\mu_Q 
\left(
\frac{\partial(\tilde T_0^2  X_2)}{\partial\mu_Q}
\right)_{T,\mu_B,\mu_S}
\nonumber\\
&\quad
+
\frac{\mu_B}{2\sqrt{\mu_Q^2+\mu_S^2}} \tilde T_0^2 
\left(
\frac{\partial X_2}{\partial\theta}
\right)_{\tilde T_0,\varphi}
-
\frac{\mu_B\mu_Q^2}{2\sqrt{\mu_Q^2+\mu_S^2}^3} \tilde T_0^2 
\left(
\frac{\partial X_2}{\partial\theta}
\right)_{\tilde T_0,\varphi}
\nonumber\\
&\quad
+
\frac{\mu_B\mu_Q}{2\sqrt{\mu_Q^2+\mu_S^2}}
\left(
\frac{\partial}{\partial\mu_Q}
\left[\tilde T_0^2 
\left(
\frac{\partial X_2}{\partial\theta}
\right)_{\tilde T_0,\varphi}
\right]
\right)_{T,\mu_B,\mu_S}
\nonumber\\
&\quad
+
\frac{\mu_S\mu_Q\mu^2}{\sqrt{\mu_Q^2+\mu_S^2}^4} \tilde T_0^2 
\left(
\frac{\partial X_2}{\partial\varphi}
\right)_{\tilde T_0,\theta}
-
\frac{\mu_Q\mu_S}{2\sqrt{\mu_Q^2+\mu_S^2}^2}
\left(
\frac{\partial X_2}{\partial\varphi}
\right)_{\tilde T_0,\theta}
\nonumber\\
&\quad
-
\frac{\mu_S\mu^2}{2\sqrt{\mu_Q^2+\mu_S^2}^2}
\left(
\frac{\partial}{\partial\mu_Q}
\left[\tilde T_0^2 
\left(
\frac{\partial X_2}{\partial\varphi}
\right)_{\tilde T_0,\theta}
\right]
\right)_{T,\mu_B,\mu_S}\bigg]\boundary.
\label{eq:A_chi2Q}
\end{align}

Similarly, differentiating the strangeness density gives
\begin{align}
T^2 \chi_2^S(T,\mu_B,\mu_Q,\mu_S)
&=
T_\text{low}^2 \chi_{2,\mathrm{HRG}}^S
+
\bigg[ \tilde T_0^2  X_2
+
\mu_S
\left(
\frac{\partial(\tilde T_0^2  X_2)}{\partial\mu_S}
\right)_{T,\mu_B,\mu_Q}
\nonumber\\
&\quad
+
\frac{\mu_B}{2\sqrt{\mu_Q^2+\mu_S^2}} \tilde T_0^2 
\left(
\frac{\partial X_2}{\partial\theta}
\right)_{\tilde T_0,\varphi}
-
\frac{\mu_B\mu_S^2}{2\sqrt{\mu_Q^2+\mu_S^2}^3} \tilde T_0^2 
\left(
\frac{\partial X_2}{\partial\theta}
\right)_{\tilde T_0,\varphi}
\nonumber\\
&\quad
+
\frac{\mu_B\mu_S}{2\sqrt{\mu_Q^2+\mu_S^2}}
\left(
\frac{\partial}{\partial\mu_S}
\left[ \tilde T_0^2 
\left(
\frac{\partial X_2}{\partial\theta}
\right)_{\tilde T_0,\varphi}
\right]
\right)_{T,\mu_B,\mu_Q}
\nonumber\\
&\quad
-
\frac{\mu_Q\mu_S\mu^2}{\sqrt{\mu_Q^2+\mu_S^2}^4} \tilde T_0^2 
\left(
\frac{\partial X_2}{\partial\varphi}
\right)_{\tilde T_0,\theta}
+
\frac{\mu_Q\mu_S}{2\sqrt{\mu_Q^2+\mu_S^2}^2} \tilde T_0^2 
\left(
\frac{\partial X_2}{\partial\varphi}
\right)_{\tilde T_0,\theta}
\nonumber\\
&\quad
+
\frac{\mu_Q\mu^2}{2\sqrt{\mu_Q^2+\mu_S^2}^2}
\left(
\frac{\partial}{\partial\mu_S}
\left[ \tilde T_0^2 
\left(
\frac{\partial X_2}{\partial\varphi}
\right)_{\tilde T_0,\theta}
\right]
\right)_{T,\mu_B,\mu_Q}\bigg]\boundary.
\label{eq:A_chi2S}
\end{align}
The chemical-potential and angular derivatives in
Eqs.~\eqref{eq:A_chi2Q} and \eqref{eq:A_chi2S} are evaluated
by the same procedure shown explicitly for
$\chi_2^B$.

\subsubsection{Second-order cross-susceptibilities}
The second-order cross-susceptibilities are calculated using the following expression:
\begin{equation}
T^2\chi_{11}^{ij}(T,\mu_B,\mu_Q,\mu_S)
=
\left(
\frac{\partial n_i}{\partial\mu_j}
\right)_{T,\mu_{k\neq j}}
\label{eq:A_cross_chi_definition}
\end{equation}

Differentiating the baryon density with respect to $\mu_Q$ gives
\begin{align}
    T^2 \chi_{11}^{BQ}(T,\mus)
    =&
    T_\text{low}^2 \chi_{11,\text{HRG}}^{BQ}
    +
    \bigg[
    \mu_B \tilde T_0^2 \left(\frac{\partial X_2}{\partial\mu_Q}\right)_{T,\mu_B,\mu_S}
    \\
    &-
    \frac{\sqrt{\mu_Q^2+\mu_S^2}}{2}
    \left(
    \frac{\partial}{\partial\mu_Q}
    \Biggl[\tilde T_0^2
    \left(
    \frac{\partial X_2}{\partial\theta}
    \right)_{\tilde T_0,\varphi}
    \Biggr]
    \right)_{T,\mu_B,\mu_S}
    \\
    &-
    \frac{\mu_Q}{2\sqrt{\mu_Q^2+\mu_S^2}}\tilde T_0^2
    \left(
    \frac{\partial X_2}{\partial\theta}
    \right)_{\tilde T_0,\varphi}
    \bigg]\boundary,
\end{align}
and, with respect to $\mu_S$,
\begin{align}
    T^2 \chi_{11}^{BS}(T,\mus)
    =&
    T_\text{low}^2 \chi_{11,\text{HRG}}^{BS}
    +
    \bigg[
    \mu_B \tilde T_0^2
    \left(
    \frac{\partial X_2}{\partial\mu_S}
    \right)_{T,\mu_B,\mu_S}
    \\
    &-
    \frac{\sqrt{\mu_Q^2+\mu_S^2}}{2}
    \left(
    \frac{\partial}{\partial\mu_S}
    \Biggl[ \tilde T_0^2
    \left(
    \frac{\partial X_2}{\partial\theta}
    \right)_{\tilde T_0,\varphi}
    \Biggr]
    \right)_{T,\mu_B,\mu_S}
    \\
    &-\frac{\mu_S}{2\sqrt{\mu_Q^2+\mu_S^2}} \tilde T_0^2
    \left(
    \frac{\partial X_2}{\partial\theta}
    \right)_{\tilde T_0,\varphi}
    \bigg]\boundary.
\end{align}

We differentiate the electric charge density with respect to $\mu_S$ to obtain
\begin{align}
    T^2 \chi_{11}^{QS}(T,\mus)
    =&
    T_\text{low}^2 \chi_{11,\text{HRG}}^{QS}
    +
    \bigg[
    \mu_Q
    \left(
    \frac{\partial(\tilde T_0^2 X_2)}{\partial\mu_S}
    \right)_{T,\mu_B,\mu_Q}
    \\
    &+
    \frac{\mu_B\mu_Q}{2\sqrt{\mu_Q^2+\mu_S^2}}
    \left(
    \frac{\partial}{\partial\mu_S}
    \Bigg[ \tilde T_0^2
    \left(
    \frac{\partial X_2}{\partial\theta}
    \right)_{\tilde T_0,\varphi}
    \Bigg]
    \right)_{T,\mu_B,\mu_Q}
    \\
    &-
    \frac{\mu_B\mu_Q\mu_S}{2\sqrt{\mu_Q^2+\mu_S^2}^3} \tilde T_0^2
    \left(
    \frac{\partial X_2}{\partial\theta}
    \right)_{\tilde T_0,\varphi}
    \\
    &-\frac{\mu_S\mu^2}{2\sqrt{\mu_Q^2+\mu_S^2}}
    \left(
    \frac{\partial}{\partial\mu_S}
    \Bigg[ \tilde T_0^2
    \left(
    \frac{\partial X_2}{\partial\varphi}
    \right)_{\tilde T_0,\theta}
    \Bigg]
    \right)_{T,\mu_B,\mu_Q}
    \\
    &-\frac{\mu^2}{2\sqrt{\mu_Q^2+\mu_S^2}^2} \tilde T_0^2
    \left(
    \frac{\partial X_2}{\partial\varphi}
    \right)_{\tilde T_0,\theta}
    -
    \frac{\mu_S^2}{\sqrt{\mu_Q^2+\mu_S^2}^2}
    \left(
    \frac{\partial X_2}{\partial\varphi}
    \right)_{\tilde T_0,\theta}
    +
    \frac{\mu_S^2\mu^2}{\sqrt{\mu_Q^2+\mu_S^2}^4}
    \left(
    \frac{\partial X_2}{\partial\varphi}
    \right)_{\tilde T_0,\theta}
    \bigg]\boundary.
\end{align}

\subsubsection{Higher-order baryon susceptibilities}

The third-order baryon susceptibility is obtained by differentiating
Eq.~\eqref{eq:A_chi2B} once more with respect to $\mu_B$,
\begin{equation}
T\chi_3^B
=
\left(
\frac{\partial (T^2 \chi_2^B)}{\partial\mu_B}
\right)_{T,\mu_Q,\mu_S}.
\label{eq:A_chi3_definition}
\end{equation}
This gives
\begin{align}
T \chi_3^B(T,\mu_B,\mu_Q,\mu_S)
&=
T\low \chi_{3,\mathrm{HRG}}^B
+
\bigg[2
\left(
\frac{\partial(\tilde T_0^2 X_2)}{\partial\mu_B}
\right)_{T,\mu_Q,\mu_S}
\nonumber\\
&\quad
+
\mu_B
\left(
\frac{\partial^2(\tilde T_0^2 X_2)}{\partial\mu_B^2}
\right)_{T,\mu_Q,\mu_S}
\nonumber\\
&\quad
-
\frac{\sqrt{\mu_Q^2+\mu_S^2}}{2}
\left(
\frac{\partial^2}{\partial\mu_B^2}
\left[ \tilde T_0^2 
\left(
\frac{\partial X_2}{\partial\theta}
\right)_{\tilde T_0,\varphi}
\right]
\right)_{T,\mu_Q,\mu_S}\bigg]\boundary.
\label{eq:A_chi3B}
\end{align}

The fourth-order baryon susceptibility follows from one additional
derivative,
\begin{equation}
\chi_4^B
=
\left(
\frac{\partial (T \chi_3^B)}{\partial\mu_B}
\right)_{T,\mu_Q,\mu_S},
\label{eq:A_chi4_definition}
\end{equation}
giving
\begin{align}
\chi_4^B(T,\mu_B,\mu_Q,\mu_S)
&=
\chi_{4,\mathrm{HRG}}^B
+
\bigg[3
\left(
\frac{\partial^2(\tilde T_0^2 X_2)}{\partial\mu_B^2}
\right)_{T,\mu_Q,\mu_S}
\nonumber\\
&\quad
+
\mu_B \tilde T_0^2 
\left(
\frac{\partial^3(\tilde T_0^2 X_2)}{\partial\mu_B^3}
\right)_{T,\mu_Q,\mu_S}
\nonumber\\
&\quad
-
\frac{\sqrt{\mu_Q^2+\mu_S^2}}{2}
\left(
\frac{\partial^3}{\partial\mu_B^3}
\left[ \tilde T_0^2 
\left(
\frac{\partial X_2}{\partial\theta}
\right)_{\tilde T_0,\varphi}
\right]
\right)_{T,\mu_Q,\mu_S}\bigg]\boundary.
\label{eq:A_chi4B}
\end{align}

\subsubsection{Higher-order electric-charge susceptibilities}

The third-order electric-charge susceptibility is obtained by differentiating
Eq.~\eqref{eq:A_chi2Q} once more with respect to $\mu_Q$,
\begin{equation}
T\chi_3^Q
=
\left(
\frac{\partial (T^2 \chi_2^Q)}{\partial\mu_Q}
\right)_{T,\mu_B,\mu_S}.
\label{eq:A_chi3Q_definition}
\end{equation}
This gives
\begin{align}
T \chi_3^Q(T,\mu_B,\mu_Q,\mu_S)
&=
T\low \chi_{3,\mathrm{HRG}}^Q
+
\bigg[
2
\left(
\frac{\partial(\tilde T_0^2 X_2)}{\partial\mu_Q}
\right)_{T,\mu_B,\mu_S}
\nonumber\\
&\quad
+
\mu_Q
\left(
\frac{\partial^2(\tilde T_0^2 X_2)}{\partial\mu_Q^2}
\right)_{T,\mu_B,\mu_S}
\nonumber\\
&\quad
+
\frac{\mu_B\mu_Q}{2\sqrt{\mu_Q^2+\mu_S^2}}
\left(
\frac{\partial^2}{\partial\mu_Q^2}
\left[
\tilde T_0^2
\left(
\frac{\partial X_2}{\partial\theta}
\right)_{\tilde T_0,\varphi}
\right]
\right)_{T,\mu_B,\mu_S}
\nonumber\\
&\quad
+
\frac{\mu_B\mu_S^2}{(\mu_Q^2+\mu_S^2)^{3/2}}
\left(
\frac{\partial}{\partial\mu_Q}
\left[
\tilde T_0^2
\left(
\frac{\partial X_2}{\partial\theta}
\right)_{\tilde T_0,\varphi}
\right]
\right)_{T,\mu_B,\mu_S}
\nonumber\\
&\quad
-
\frac{3\mu_B\mu_Q\mu_S^2}{2(\mu_Q^2+\mu_S^2)^{5/2}}
\tilde T_0^2
\left(
\frac{\partial X_2}{\partial\theta}
\right)_{\tilde T_0,\varphi}
\nonumber\\
&\quad
-
\frac{\mu_S\mu^2}{2(\mu_Q^2+\mu_S^2)}
\left(
\frac{\partial^2}{\partial\mu_Q^2}
\left[
\tilde T_0^2
\left(
\frac{\partial X_2}{\partial\varphi}
\right)_{\tilde T_0,\theta}
\right]
\right)_{T,\mu_B,\mu_S}
\nonumber\\
&\quad
+
\frac{2\mu_B^2\mu_Q\mu_S}{(\mu_Q^2+\mu_S^2)^2}
\left(
\frac{\partial}{\partial\mu_Q}
\left[
\tilde T_0^2
\left(
\frac{\partial X_2}{\partial\varphi}
\right)_{\tilde T_0,\theta}
\right]
\right)_{T,\mu_B,\mu_S}
\nonumber\\
&\quad
-
\frac{\mu_B^2\mu_S(3\mu_Q^2-\mu_S^2)}
     {(\mu_Q^2+\mu_S^2)^3}
\tilde T_0^2
\left(
\frac{\partial X_2}{\partial\varphi}
\right)_{\tilde T_0,\theta}
\bigg]\boundary.
\label{eq:A_chi3Q}
\end{align}

The fourth-order electric-charge susceptibility follows from one additional
derivative,
\begin{equation}
\chi_4^Q
=
\left(
\frac{\partial (T \chi_3^Q)}{\partial\mu_Q}
\right)_{T,\mu_B,\mu_S},
\label{eq:A_chi4Q_definition}
\end{equation}
giving
\begin{align}
\chi_4^Q(T,\mu_B,\mu_Q,\mu_S)
&=
\chi_{4,\mathrm{HRG}}^Q
+
\bigg[
3
\left(
\frac{\partial^2(\tilde T_0^2 X_2)}{\partial\mu_Q^2}
\right)_{T,\mu_B,\mu_S}
\nonumber\\
&\quad
+
\mu_Q
\left(
\frac{\partial^3(\tilde T_0^2 X_2)}{\partial\mu_Q^3}
\right)_{T,\mu_B,\mu_S}
\nonumber\\
&\quad
+
\frac{\mu_B\mu_Q}{2\sqrt{\mu_Q^2+\mu_S^2}}
\left(
\frac{\partial^3}{\partial\mu_Q^3}
\left[
\tilde T_0^2
\left(
\frac{\partial X_2}{\partial\theta}
\right)_{\tilde T_0,\varphi}
\right]
\right)_{T,\mu_B,\mu_S}
\nonumber\\
&\quad
+
\frac{3\mu_B\mu_S^2}{2(\mu_Q^2+\mu_S^2)^{3/2}}
\left(
\frac{\partial^2}{\partial\mu_Q^2}
\left[
\tilde T_0^2
\left(
\frac{\partial X_2}{\partial\theta}
\right)_{\tilde T_0,\varphi}
\right]
\right)_{T,\mu_B,\mu_S}
\nonumber\\
&\quad
-
\frac{9\mu_B\mu_Q\mu_S^2}{2(\mu_Q^2+\mu_S^2)^{5/2}}
\left(
\frac{\partial}{\partial\mu_Q}
\left[
\tilde T_0^2
\left(
\frac{\partial X_2}{\partial\theta}
\right)_{\tilde T_0,\varphi}
\right]
\right)_{T,\mu_B,\mu_S}
\nonumber\\
&\quad
+
\frac{3\mu_B\mu_S^2(4\mu_Q^2-\mu_S^2)}
     {2(\mu_Q^2+\mu_S^2)^{7/2}}
\tilde T_0^2
\left(
\frac{\partial X_2}{\partial\theta}
\right)_{\tilde T_0,\varphi}
\nonumber\\
&\quad
-
\frac{\mu_S\mu^2}{2(\mu_Q^2+\mu_S^2)}
\left(
\frac{\partial^3}{\partial\mu_Q^3}
\left[
\tilde T_0^2
\left(
\frac{\partial X_2}{\partial\varphi}
\right)_{\tilde T_0,\theta}
\right]
\right)_{T,\mu_B,\mu_S}
\nonumber\\
&\quad
+
\frac{3\mu_B^2\mu_Q\mu_S}{(\mu_Q^2+\mu_S^2)^2}
\left(
\frac{\partial^2}{\partial\mu_Q^2}
\left[
\tilde T_0^2
\left(
\frac{\partial X_2}{\partial\varphi}
\right)_{\tilde T_0,\theta}
\right]
\right)_{T,\mu_B,\mu_S}
\nonumber\\
&\quad
-
\frac{3\mu_B^2\mu_S(3\mu_Q^2-\mu_S^2)}
     {(\mu_Q^2+\mu_S^2)^3}
\left(
\frac{\partial}{\partial\mu_Q}
\left[
\tilde T_0^2
\left(
\frac{\partial X_2}{\partial\varphi}
\right)_{\tilde T_0,\theta}
\right]
\right)_{T,\mu_B,\mu_S}
\nonumber\\
&\quad
+
\frac{12\mu_B^2\mu_Q\mu_S(\mu_Q^2-\mu_S^2)}
     {(\mu_Q^2+\mu_S^2)^4}
\tilde T_0^2
\left(
\frac{\partial X_2}{\partial\varphi}
\right)_{\tilde T_0,\theta}
\bigg]\boundary.
\label{eq:A_chi4Q}
\end{align}

\subsubsection{Higher-order strangeness susceptibilities}

The third-order strangeness susceptibility is obtained by differentiating
Eq.~\eqref{eq:A_chi2S} once more with respect to $\mu_S$,
\begin{equation}
\chi_3^S
=
\left(
\frac{\partial(T^2 \chi_2^S)}{\partial\mu_S}
\right)_{T,\mu_B,\mu_Q}.
\label{eq:A_chi3S_definition}
\end{equation}
This gives
\begin{align}
T \chi_3^S(T,\mu_B,\mu_Q,\mu_S)
&=
T\low \chi_{3,\mathrm{HRG}}^S
+
\bigg[
2
\left(
\frac{\partial(\tilde T_0^2 X_2)}{\partial\mu_S}
\right)_{T,\mu_B,\mu_Q}
\nonumber\\
&\quad
+
\mu_S
\left(
\frac{\partial^2(\tilde T_0^2 X_2)}{\partial\mu_S^2}
\right)_{T,\mu_B,\mu_Q}
\nonumber\\
&\quad
+
\frac{\mu_B\mu_S}{2\sqrt{\mu_Q^2+\mu_S^2}}
\left(
\frac{\partial^2}{\partial\mu_S^2}
\left[
\tilde T_0^2
\left(
\frac{\partial X_2}{\partial\theta}
\right)_{\tilde T_0,\varphi}
\right]
\right)_{T,\mu_B,\mu_Q}
\nonumber\\
&\quad
+
\frac{\mu_B\mu_Q^2}{(\mu_Q^2+\mu_S^2)^{3/2}}
\left(
\frac{\partial}{\partial\mu_S}
\left[
\tilde T_0^2
\left(
\frac{\partial X_2}{\partial\theta}
\right)_{\tilde T_0,\varphi}
\right]
\right)_{T,\mu_B,\mu_Q}
\nonumber\\
&\quad
-
\frac{3\mu_B\mu_S\mu_Q^2}{2(\mu_Q^2+\mu_S^2)^{5/2}}
\tilde T_0^2
\left(
\frac{\partial X_2}{\partial\theta}
\right)_{\tilde T_0,\varphi}
\nonumber\\
&\quad
+
\frac{\mu_Q\mu^2}{2(\mu_Q^2+\mu_S^2)}
\left(
\frac{\partial^2}{\partial\mu_S^2}
\left[
\tilde T_0^2
\left(
\frac{\partial X_2}{\partial\varphi}
\right)_{\tilde T_0,\theta}
\right]
\right)_{T,\mu_B,\mu_Q}
\nonumber\\
&\quad
-
\frac{2\mu_B^2\mu_S\mu_Q}{(\mu_Q^2+\mu_S^2)^2}
\left(
\frac{\partial}{\partial\mu_S}
\left[
\tilde T_0^2
\left(
\frac{\partial X_2}{\partial\varphi}
\right)_{\tilde T_0,\theta}
\right]
\right)_{T,\mu_B,\mu_Q}
\nonumber\\
&\quad
+
\frac{\mu_B^2\mu_Q(3\mu_S^2-\mu_Q^2)}
     {(\mu_Q^2+\mu_S^2)^3}
\tilde T_0^2
\left(
\frac{\partial X_2}{\partial\varphi}
\right)_{\tilde T_0,\theta}
\bigg]\boundary.
\label{eq:A_chi3S}
\end{align}

The fourth-order strangeness susceptibility follows from one additional
derivative,
\begin{equation}
\chi_4^S
=
\left(
\frac{\partial(T \chi_3^S)}{\partial\mu_S}
\right)_{T,\mu_B,\mu_Q},
\label{eq:A_chi4S_definition}
\end{equation}
giving
\begin{align}
\chi_4^S(T,\mu_B,\mu_Q,\mu_S)
&=
\chi_{4,\mathrm{HRG}}^S
+
\bigg[
3
\left(
\frac{\partial^2(\tilde T_0^2 X_2)}{\partial\mu_S^2}
\right)_{T,\mu_B,\mu_Q}
\nonumber\\
&\quad
+
\mu_S
\left(
\frac{\partial^3(\tilde T_0^2 X_2)}{\partial\mu_S^3}
\right)_{T,\mu_B,\mu_Q}
\nonumber\\
&\quad
+
\frac{\mu_B\mu_S}{2\sqrt{\mu_Q^2+\mu_S^2}}
\left(
\frac{\partial^3}{\partial\mu_S^3}
\left[
\tilde T_0^2
\left(
\frac{\partial X_2}{\partial\theta}
\right)_{\tilde T_0,\varphi}
\right]
\right)_{T,\mu_B,\mu_Q}
\nonumber\\
&\quad
+
\frac{3\mu_B\mu_Q^2}{2(\mu_Q^2+\mu_S^2)^{3/2}}
\left(
\frac{\partial^2}{\partial\mu_S^2}
\left[
\tilde T_0^2
\left(
\frac{\partial X_2}{\partial\theta}
\right)_{\tilde T_0,\varphi}
\right]
\right)_{T,\mu_B,\mu_Q}
\nonumber\\
&\quad
-
\frac{9\mu_B\mu_S\mu_Q^2}{2(\mu_Q^2+\mu_S^2)^{5/2}}
\left(
\frac{\partial}{\partial\mu_S}
\left[
\tilde T_0^2
\left(
\frac{\partial X_2}{\partial\theta}
\right)_{\tilde T_0,\varphi}
\right]
\right)_{T,\mu_B,\mu_Q}
\nonumber\\
&\quad
+
\frac{3\mu_B\mu_Q^2(4\mu_S^2-\mu_Q^2)}
     {2(\mu_Q^2+\mu_S^2)^{7/2}}
\tilde T_0^2
\left(
\frac{\partial X_2}{\partial\theta}
\right)_{\tilde T_0,\varphi}
\nonumber\\
&\quad
+
\frac{\mu_Q\mu^2}{2(\mu_Q^2+\mu_S^2)}
\left(
\frac{\partial^3}{\partial\mu_S^3}
\left[
\tilde T_0^2
\left(
\frac{\partial X_2}{\partial\varphi}
\right)_{\tilde T_0,\theta}
\right]
\right)_{T,\mu_B,\mu_Q}
\nonumber\\
&\quad
-
\frac{3\mu_B^2\mu_S\mu_Q}{(\mu_Q^2+\mu_S^2)^2}
\left(
\frac{\partial^2}{\partial\mu_S^2}
\left[
\tilde T_0^2
\left(
\frac{\partial X_2}{\partial\varphi}
\right)_{\tilde T_0,\theta}
\right]
\right)_{T,\mu_B,\mu_Q}
\nonumber\\
&\quad
+
\frac{3\mu_B^2\mu_Q(3\mu_S^2-\mu_Q^2)}
     {(\mu_Q^2+\mu_S^2)^3}
\left(
\frac{\partial}{\partial\mu_S}
\left[
\tilde T_0^2
\left(
\frac{\partial X_2}{\partial\varphi}
\right)_{\tilde T_0,\theta}
\right]
\right)_{T,\mu_B,\mu_Q}
\nonumber\\
&\quad
+
\frac{12\mu_B^2\mu_Q\mu_S(\mu_Q^2-\mu_S^2)}
     {(\mu_Q^2+\mu_S^2)^4}
\tilde T_0^2
\left(
\frac{\partial X_2}{\partial\varphi}
\right)_{\tilde T_0,\theta}
\bigg]\boundary.
\label{eq:A_chi4S}
\end{align}

\section{Specific heat and sound velocity in multi-component systems}
\label{app:hessian}

Specific heat is proportional to the entropy density temperature derivative taken while conserved charge densities are held constant, e.g. in (2+1)-flavor QCD one has
\begin{align}
    c_V=T\left(\frac{\partial s}{\partial T}\right)_{n_B,n_Q,n_S}.
\end{align}
Similarly, the adiabatic speed of sound squared quantifies the pressure change with energy density along constant ratios of conserved charge densities over the entropy density,
\begin{align}
    c_s^2=\left(\frac{\partial P}{\partial\epsilon}\right)_{n_B/s,n_Q/s,n_S/s}.
\end{align}

In the grand-canonical ensemble, we compute the primary quantities while holding chemical potentials fixed, rather than charge densities.
It is therefore useful to express $c_V$ and $c_s^2$ in terms of grand-canonical ensemble quantities.
To our knowledge, such expressions have been written for a single conserved charge~\cite{Floerchinger:2015efa}, but no compact expression has been given for a general multi-dimensional case.
Here we provide such expressions valid for arbitrary multi-component system, including the $B$, $Q$, $S$ charges studied here.

For the following derivations, we assume a generic system with $f$ independent conserved charges and we use the following vector notation
\begin{align}
    &\mathbf{x}\equiv(s,\mathbf{n}), &\mathbf{y} \equiv(T,\boldsymbol{\mu}), &&\hat{\chi} \equiv \chi_{ij}=\left(\frac{\partial^2 (P/T^4)}{\partial(\mu_i/T)\partial(\mu_j/T)}\right)_T.
\end{align}
Here $\mathbf{n}$ and $\boldsymbol{\mu}$ are the vectors of conserved charge densities and the corresponding chemical potentials.
For the case of (2+1)-flavor QCD in the $(B,Q,S)$ basis, one has $\mathbf{n} = (n_B,n_Q,n_S)$ and $\boldsymbol{\mu} = (\mu_B, \mu_Q, \mu_S)$.
We note that $\mathbf{x}$ and $\mathbf{y}$ are connected through the following thermodynamic relation
\begin{align}
\label{eq:TDxy}
\mathbf{x} = \frac{\partial P(\mathbf{y})}{\partial \mathbf{y}}.
\end{align}

We write the pressure Hessian as
\begin{align}
    \hat H=\frac{\partial^2 P}{\partial(T,\boldsymbol{\mu})^2}=
    \begin{pmatrix}
        s_T & \mathbf{b}^T \\
        \mathbf{b} & \hat{A}
    \end{pmatrix},
\end{align}
where $s_T=(\partial s/\partial T)_{\mu}$, $\mathbf{b}\equiv(\partial\mathbf{n}/\partial T)_{\mu_B,\mu_Q,\mu_S}$, and $\hat{A}=T^2\hat{\chi}$ following the thermodynamic relations.

\subsubsection{Specific heat}

The specific heat is defined as
\begin{align}
    c_V=T\left(\frac{\partial s}{\partial T}\right)_{\mathbf{n}}.
\end{align}
% which we would like to express in terms of the Hessian.
% 
Using $\mathbf{x}=\mathbf{x}(\mathbf{y})$, we have
\begin{align}
    d\mathbf{x}=\hat{H} d\mathbf{y} \quad \Rightarrow \quad d\mathbf{y}=\hat{H}^{-1}d\mathbf{x},
\end{align}
which allows us to write
\begin{align}
    dT=\sum_{j=0}^fH^{-1}_{0j}dx_j=H^{-1}_{00}ds + \sum_{j=1}^f H^{-1}_{0j}dn_j,
\end{align}
giving
\begin{align}
    \left(\frac{\partial T}{\partial s}\right)_{\mathbf{n}}=H^{-1}_{00}
    \qquad \Rightarrow \qquad
    \left(\frac{\partial s}{\partial T}\right)_{\mathbf{n}}=\frac{1}{H^{-1}_{00}}.
\end{align}

An inverse matrix element can be expressed in terms of its cofactor.
For the $(0,0)$ element, the cofactor is the determinant of the matrix $\hat{A}$.
One obtains,
\begin{align}
    H^{-1}_{00}=\frac{\det\hat{A}}{\det \hat H},
\end{align}
yielding a compact matrix expression for the specific heat
\begin{align}
    c_V=T\frac{\det \hat H}{\det\hat{A}}.
\end{align}

\subsubsection{The speed of sound squared}

The squared speed of sound is defined as
\begin{align}
    c_s^2=\left(\frac{\partial P}{\partial\epsilon}\right)_{\mathbf{n}/s}.
\end{align}

For constant $\mathbf{n}/s$, one has  $d \mathbf{n}=\mathbf{n}/s~ds$. 
We choose the entropy density as the independent variable under constraints $d\mathbf{n}=\mathbf{n}/s~ds$, which allows us to write
\begin{align}
    c_s^2=\frac{(\partial P/\partial s)_{\mathbf{n}/s}}{(\partial\epsilon/\partial s)_{\mathbf{n}/s}}.
\end{align}

Considering the pressure $P = P(\mathbf{x}) = P(s, \mathbf{n})$, one can write
\begin{align}
dP & = (\partial P / \partial s)_{\mathbf{n}} ds + \sum_{j=1}^f (\partial P / \partial n_j)_{s,n_{i \neq j}} dn_j.
\end{align}
From $d \mathbf{n}=\mathbf{n}/s~ds$ one then obtains
\begin{align}
\left(\frac{\partial P}{\partial s}\right)_{\mathbf{n}/s} = \left(\frac{\partial P}{\partial s}\right)_{\mathbf{n}}+ \left(\frac{\partial P}{\partial \mathbf{n}}\right)_{s} \frac{\mathbf{n}}{s}.
\end{align}

For the energy density, $\epsilon = \epsilon(\mathbf{x}) = \epsilon(s, \mathbf{n})$, one can use the Gibbs relation
\begin{align}
d\epsilon & = T ds + \boldsymbol{\mu} \cdot d \mathbf{n}.
\end{align}
Together with $d \mathbf{n}=\mathbf{n}/s~ds$ this gives
\begin{align}
\left(\frac{\partial \epsilon}{\partial s}\right)_{\mathbf{n}/s} & = T + \frac{\boldsymbol{\mu} \cdot \mathbf{n}}{s} \nonumber \\
& = \frac{\epsilon + P}{s}~.
\end{align}

We now have
\begin{align}
    c_s^2 & =\frac{\left(\frac{\partial P}{\partial s}\right)_{\mathbf{n}} s+ \left(\frac{\partial P}{\partial \mathbf{n}}\right)_{s} \mathbf{n}}{\epsilon + P} \nonumber \\
    & = \frac{\left(\frac{\partial P}{\partial \mathbf{x}}\right)  \mathbf{x}}{\epsilon + P}.
\end{align}

To express the speed of sound squared in terms of the Hessian, we first expand the derivative in the numerator as
\begin{align}
    \left(\frac{\partial P}{\partial x_i}\right)_{x_{j\neq i}}&=\sum_{j=0}^f\left(\frac{\partial P}{\partial y_j}\right)_{y_{k\neq j}}\left(\frac{\partial y_j}{\partial x_i}\right)_{x_{k\neq i}} \nonumber
    \\
    &=\sum_{j=0}^2x_j \hat H_{ji}^{-1},
\end{align}
where we have used Eq.~\eqref{eq:TDxy} and the definition of the Hessian $\hat H_{ij}^{-1}=(\partial y_i/\partial x_j)_{x_{k\neq j}}$.
This gives the final expression for the sound velocity squared in matrix form:
\begin{align}
    c_s^2=\frac{\sum_{i,j=0}^fx_i\hat H_{ij}^{-1}x_j}{\epsilon+P}.
\end{align}

\end{appendix}

\end{document}